\documentclass[prb,aps,twocolumn]{revtex4-1}
\usepackage{graphics, bm, psfrag, amsmath, amssymb, epsfig, grffile, float, bbold}
\usepackage{subfigure}
\usepackage{dsfont}
\usepackage{color}
\usepackage{hyperref}
\newcommand*{\citen}[1]{%
  \begingroup
    \romannumeral-`\x 
    \setcitestyle{numbers}%
    \cite{#1}%
  \endgroup   
}

\begin{document}

\title{Odd-frequency pairing in a superconductor coupled to two parallel nanowires}

\author{Christopher Triola}
\affiliation{Department of Physics and Astronomy, Uppsala University, Box 516, S-751 20 Uppsala, Sweden}
\author{Annica M. Black-Schaffer}
\affiliation{Department of Physics and Astronomy, Uppsala University, Box 516, S-751 20 Uppsala, Sweden}

\begin{abstract}

We study the behavior of Cooper pair amplitudes that emerge when a two-dimensional superconductor is coupled to two parallel nanowires, focusing on the conditions for realizing odd-frequency pair amplitudes in the absence of spin-orbit coupling or magnetism. In general, any finite tunneling between the superconductor and the two nanowires induces odd-frequency spin-singlet pair amplitudes in the substrate as well as a substantial odd-frequency interwire pairing, both of which vanish locally. Interestingly, in the regime of strong superconductor-nanowire tunneling, we find that the presence of two nanowires allows for the conversion of non-local odd-frequency pairing to local even-frequency pairing. By studying this higher-order symmetry conversion process, we are able to identify a notable effect of the odd-frequency pairing in the superconductor on local quantities accessible by experiments. Specifically, we find that the odd-frequency pairing plays a direct role in the emergence of certain subgap features in the local density of states, and, importantly, it is responsible for a reduction of the maximum Josephson current between the two nanowires, measurable using Josephson scanning tunneling microscopy. We discuss ways to control the sizes of these effects induced by odd-frequency superconductivity by tuning the parameters describing the nanowires.

\end{abstract}


\maketitle

\section{Introduction}

The study of proximity-induced superconductivity in one-dimensional (1D) nanowires has generated a great deal of interest in recent years, driven primarily by their potential for realizing states with non-Abelian statistics holding the promise for topological quantum computation.\cite{alicea2012new, SternLindner2013, SarmaNPG2015} The simplest proposals involve single nanowires with Rashba spin-orbit coupling in proximity to a conventional $s$-wave superconductor and in the presence of an applied magnetic field.\cite{lutchyn2010majorana,oreg2010helical} However, the non-Abelian Majorana bound states found in these single nanowire systems are Ising anyons and cannot be used to construct all gates necessary for universal quantum computation\cite{trebst2008short} in contrast to e.g.~Fibonacci anyons.\cite{trebst2008short,mong2014universal,hu2018fibonacci} Fibonacci anyons can be created using parafermions, exotic excitations which can be realized by coupling two nanowires with Rashba spin-orbit coupling to an $s$-wave superconductor, in the absence of a magnetic field.\cite{klinovaja2014time,gaidamauskas2014majorana,ebisu2016theory,thakurathi2018majorana} 

An important feature of double nanowire systems, not present in single nanowires, is the possibility for crossed Andreev reflection processes, which connects the superconducting pairs in the two wires, and is known to play a significant role in the physics of these systems.\cite{reeg2017destructive} Given that regular Andreev reflection processes have been shown to be related to the generation of Cooper pairs with unconventional symmetries in single wire systems,\cite{cayao2017odd,keidel2018tunable,cayao2018odd} it is interesting to consider the different pair symmetries that can arise in double nanowire systems. Moreover, the potential use of double nanowires in new technologies, with experiments already working to characterize their properties,\cite{baba2018cooper} further highlights the importance of a theoretical analysis of the pairing and symmetries exhibited by these systems. Such an analysis will not only provide a deeper understanding of the electronic properties of these systems but also potentially suggest new ways to utilize them for practical applications.

It is well-established that the fermionic nature of electrons tightly constrains the allowed symmetries of the Cooper pairs and thus the superconducting gap function. Specifically, in the limit of equal-time pairing and a single-component gap, spatially even-parity gap functions (like $s$- or $d$-wave) must correspond to spin-singlet states, while odd-parity gap functions ($p$- or $f$-wave) must correspond to spin-triplet states. However, if the electrons comprising the condensate are paired at unequal times the superconducting gap can also be odd in time or, equivalently, odd in frequency (odd-$\omega$), allowing the condensate to be even in spatial parity and spin-triplet or odd-parity and spin-singlet\cite{linder2017odd}. This possibility, originally posited for $^3$He by Berezinskii\cite{Berezinskii1974} and then later for superconductivity,\cite{kirkpatrick_1991_prl,belitz_1992_prb,BalatskyPRB1992} is intriguing both because of the unconventional symmetries which it permits and for the fact that it represents a class of hidden order, due to the vanishing of equal-time correlations. 

While the thermodynamic stability of intrinsically odd-$\omega$ phases has, so far, only been discussed as a theoretical possibility,\cite{coleman_1993_prl,coleman_1994_prb,coleman_1995_prl,heid1995thermodynamic,belitz_1999_prb,solenov2009thermodynamical,kusunose2011puzzle,FominovPRB2015} significant progress has been made understanding the way in which odd-$\omega$ pairing can be induced by altering a system's conventional superconducting correlations.\cite{BergeretPRL2001, bergeret2005odd, halterman2007odd, yokoyama2007manifestation, houzet2008ferromagnetic, EschrigNat2008, LinderPRB2008, crepin2015odd, YokoyamaPRB2012, Black-SchafferPRB2012, Black-SchafferPRB2013, TriolaPRB2014, tanaka2007theory, TanakaPRB2007, LinderPRL2009, LinderPRB2010_2, TanakaJPSJ2012, triola2016prl, triolaprb2016, black2013odd, sothmann2014unconventional, parhizgar_2014_prb, asano2015odd, komendova2015experimentally, burset2016all, komendova2017odd, kuzmanovski2017multiple, triola2017pair,keidel2018tunable, triola2018odd,fleckenstein2018conductance,asano2018green} The best established example is found in ferromagnet-superconductor junctions,\cite{BergeretPRL2001, bergeret2005odd, halterman2007odd, yokoyama2007manifestation, houzet2008ferromagnetic, EschrigNat2008, LinderPRB2008, crepin2015odd} in which experiments have observed key signatures of odd-$\omega$ spin-triplet pair correlations,\cite{zhu2010angular,di2015signature,di2015intrinsic} despite using conventional spin-singlet $s$-wave superconductors. Additionally, odd-parity odd-$\omega$ pair amplitudes been shown to be ubiquitous at interfaces between normal metals (N) and conventional spin-singlet superconductors (S), with close connections to observed McMillan-Rowell oscillations, as well as midgap Andreev resonances.\cite{tanaka2007theory, TanakaPRB2007} For a modern review of odd-$\omega$ superconductivity, see Ref.~[\citen{linder2017odd}].

Inspired by the previous works on S/N interfaces, in this work we consider a seemingly related, but yet quite different, system: two parallel nanowires coupled to a superconducting substrate with conventional spin-singlet $s$-wave order parameter, as shown in Fig. \ref{fig:setup}. We examine the symmetries of the emergent Cooper pair amplitudes, focusing on the appearance of odd-$\omega$ superconductivity and its physical consequences. In particular, by expanding the anomalous Green's function to leading order in the superconductor-nanowire tunneling parameters, we find that odd-$\omega$ spin-singlet odd-parity Cooper pair amplitudes emerge in both the substrate and the interwire channel. The appearance of the odd-$\omega$ pair amplitudes in the substrate are consistent with the above-mentioned works on S/N junctions,\cite{tanaka2007theory, TanakaPRB2007} while the pair amplitudes in the wires are consistent with previous results modeling Rashba quantum wires,\cite{ebisu2016theory} as well as analogous multiterminal/multiband systems \cite{black2013odd, sothmann2014unconventional, parhizgar_2014_prb,  asano2015odd, komendova2015experimentally, burset2016all, komendova2017odd, kuzmanovski2017multiple, triola2017pair,keidel2018tunable, triola2018odd,fleckenstein2018conductance,asano2018green}. However, in contrast to previous studies, we consider how the odd-$\omega$ pairing induced by one of the nanowires is affected by the presence of the other nanowire, an effect that shows up as higher-order cross-terms in the diagrammatic expansion of the pair amplitudes. Importantly, we then find that these higher-order processes lead to a conversion of odd-$\omega$ odd-parity amplitudes to even-$\omega$ even-parity amplitudes. Since these reconverted even-parity amplitudes are local in space, this reconversion process allows the original odd-$\omega$ odd-parity Cooper pairs to have a direct and measurable impact on easily measurable \textit{local} observables.

More specifically, we derive explicit expressions characterizing the symmetry conversion of odd-parity odd-$\omega$ pairing to local even-$\omega$ pair amplitudes, to infinite order in the superconductor-nanowire tunneling. This analysis establishes that two nanowires are needed for this process. We then study the features generated by the odd-$\omega$ pairing in two local and highly accessible experimental observables: the local density of states (LDOS), measurable by scanning tunneling microscopy (STM), and local Josephson current (LJC), measurable by Josephson STM.\cite{vsmakov2001josephson, naaman2001fluctuation,kashuba2017majorana} In the LDOS, we show that large subgap peaks emerge due to the presence of the nanowires, and we find that one of these spectral features is directly related to higher-order symmetry conversion of original odd-$\omega$ pairing. In the LJC, we find that the odd-$\omega$ pairing contributes directly to a noticeable reduction of the LJC maximum value in the region between the two nanowires, and we show how this reduction can be tuned by adjusting different physical parameters. These results establish both that odd-$\omega$ superconductivity is generated in double nanowire systems and that, despite its non-local nature, odd-$\omega$ superconductivity has profound effects on easily measurable local quantities. 

The remainder of this work is organized as follows. In Sec.~\ref{sec:mod} we introduce the model we will use to study the double nanowire-superconductor system and define the Green's functions used throughout this work. In Sec.~\ref{sec:odd} we derive the perturbative corrections to the anomalous Green's functions and establish the existence of odd-$\omega$ pairing in the presence of finite nanowire-superconductor tunneling. In Sec.~\ref{sec:high} we examine the higher-order corrections to the Green's functions and show that it is exactly the presence of two nanowires that allows the conversion of odd-$\omega$ amplitudes to even-$\omega$ amplitudes with novel properties. In Sec.~\ref{sec:obs} we study the effect on local observables of the odd-$\omega$ pairing through the higher-order symmetry conversions, identifying clearly measurable features in both LDOS and LJC. Finally, in Sec.~\ref{sec:conclusions} we conclude our work.

\section{Model}
\label{sec:mod}

\begin{figure}
 \begin{center}
  \centering
        \includegraphics[width=0.4\textwidth]{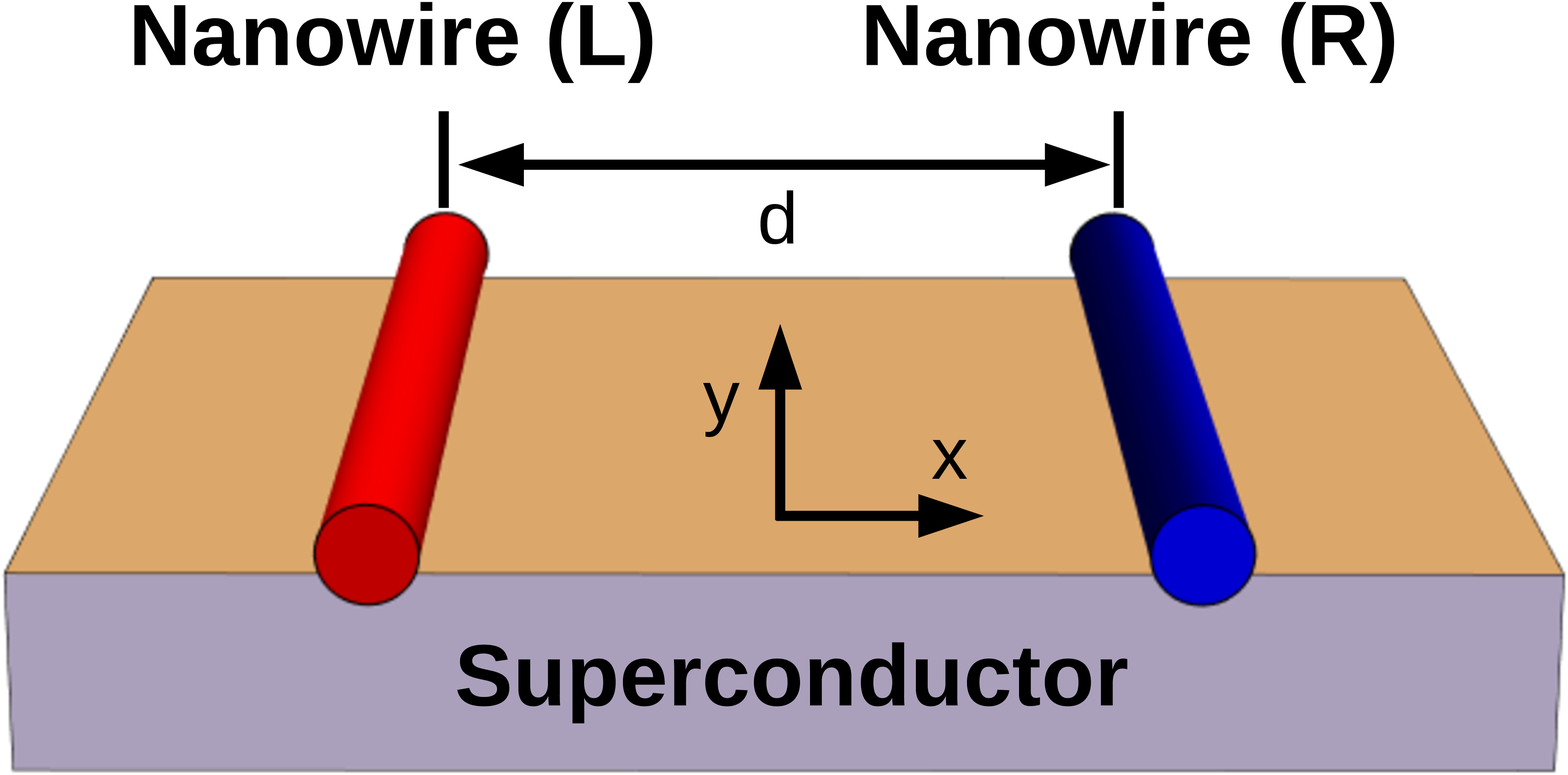}
  \caption{Schematic of the system described by the Hamiltonian in Eq. (\ref{eq:hamilton}): two parallel nanowires, separated by a distance $d$, coupled to the 2D surface of a conventional spin-singlet $s$-wave superconductor. Throughout this work the nanowires are assumed to be infinitely long 1D objects in the $y$-direction and the superconductor is assumed span in the entire 2D plane.}
  \label{fig:setup}
 \end{center}
\end{figure}

We wish to study the emergent symmetries of superconductivity in a physical system composed of two parallel nanowires separated by a distance $d$ coupled to a conventional superconducting substrate, which we model as a two-dimensional (2D) spin-singlet $s$-wave superconductor, see Fig. \ref{fig:setup}. Throughout this work we assume the nanowires are non-magnetic and possess no appreciable spin-orbit coupling, such that the system has only trivial spin structure. This eliminates the possibility of realizing odd-$\omega$, spin-triplet, even-parity (i.e.~local) superconducting pairing, and thus the only odd-$\omega$ pairing allowed in this system is of the spin-singlet, odd-parity (i.e. non-local) type.
To capture the essential physics of this system we employ a Hamiltonian of the form $H=H^{\text{L}}_{\text{NW}}+H^{\text{R}}_{\text{NW}}+H_{\text{SC}}+H^{\text{L}}_{\text{t}}+H^{\text{R}}_{\text{t}}$ where
\begin{widetext}
\begin{equation}
\begin{aligned}
H_{\text{SC}}&=\sum_\sigma\int dx \frac{dk_y}{2\pi}  d^\dagger_{x,k_y,\sigma}\left[-\frac{\partial_x^2}{2m_s}+\xi_{s,k_y}\right]d_{x,k_y,\sigma}+\int  dx \frac{dk_y}{2\pi}  \Delta d^\dagger_{x,-k_y,\uparrow}d^\dagger_{x,k_y,\downarrow} + \text{H.c.}, \\
H^{\text{i}}_{\text{NW}}&=\sum_\sigma\int \frac{dk_y}{2\pi} \xi_{i,k_y} c^\dagger_{i,k_y,\sigma}c_{i,k_y,\sigma}, \\
H^{i}_{\text{t}}&=-t_i\sum_\sigma\int \frac{dk_y}{2\pi} c^\dagger_{i,k_y,\sigma}d_{x_i,k_y,\sigma} + \text{H.c.},
\end{aligned}
\label{eq:hamilton}
\end{equation}
\end{widetext}
where $d^\dagger_{x,k_y,\sigma}$ ($d_{x,k_y,\sigma}$) creates (annihilates) a quasiparticle state in the superconducting substrate with spin $\sigma$ at position $x$ along the axis perpendicular to the nanowires and with momentum $k_y$ along the axis parallel to the nanowires. Likewise, $c^\dagger_{i,k_y,\sigma}$ ($c_{i,k_y,\sigma}$) creates (annihilates) a quasiparticle state with spin $\sigma$ and momentum $k_y$ in nanowire $i=L,R$. Moreover, $\xi_{s,k_y}=\tfrac{k_y^2}{2m_s}-\mu_s$ is the normal state quasiparticle dispersion in the superconductor along the $y$-axis set by the effective mass $m_s$ and measured from the chemical potential $\mu_s$, while $\xi_{i,k_y}=\tfrac{k_y^2}{2m_i}-\mu_i$ is the quasiparticle dispersion in the $i^{{\text{th}}}$ nanowire set by the effective mass $m_i$ and measured from the chemical potential $\mu_i$. The superconducting substrate is further described by the order parameter $\Delta$ which has spin-singlet $s$-wave symmetry. Finally, $t_i$ is the tunneling amplitude coupling the superconductor to the $i^{\text{th}}$ nanowire, which is located at position $x=x_i$. 

To study the emergent electronic properties of the system described by the Hamiltonian in Eq.~(\ref{eq:hamilton}), we begin by noting that the presence of the nanowires breaks translation-invariance along the $x$-direction, but not the $y$-direction, thus allowing us to keep a $k_y$ reciprocal coordinate. As a consequence, we define the normal and anomalous Green's functions for the superconductor as:
\begin{equation}
\begin{aligned}
G^{s}_{x_1,x_2;k_y;\tau}&=-\langle T_\tau d_{x_1,k_y,\uparrow}(\tau)d^\dagger_{x_2,k_y,\uparrow}(0) \rangle, \\
F^{s}_{x_1,x_2;k_y;\tau}&=-\langle T_\tau d_{x_1,k_y,\uparrow}(\tau)d_{x_2,-k_y,\downarrow}(0) \rangle, 
\end{aligned}
\label{eq:g_f_sc}
\end{equation}
where $\tau$ is an imaginary time, and $T_\tau$ is the usual $\tau$-ordering operator for fermions. Similarly, we define the normal and anomalous Green's functions for the electronic excitations in the two nanowires as:
\begin{equation}
\begin{aligned}
G^{n}_{i,j;k_y;\tau}&=-\langle T_\tau c_{i,k_y,\uparrow}(\tau)c^\dagger_{j,k_y,\uparrow}(0) \rangle, \\
F^{n}_{i,j;k_y;\tau}&=-\langle T_\tau c_{i,k_y,\uparrow}(\tau)c_{j,-k_y,\downarrow}(0) \rangle, 
\end{aligned}
\label{eq:g_f_n}
\end{equation}
where, due to coupling through the superconducting substrate, we allow for both intrawire ($i=j$) and interwire ($i\neq j$) correlations in Eq.~(\ref{eq:g_f_n}).

As usual, we find it convenient to combine the normal and anomalous Green's functions into the following Nambu space Green's functions for the superconductor and the nanowires:
\begin{equation}
\begin{aligned}
\hat{\mathcal{G}}_s(x_1,x_2;k_y;i\omega_n)&=\left(
\begin{array}{cc}
G^{s}_{x_1,x_2;k_y;i\omega_n} & F^{s}_{x_1,x_2;k_y;i\omega_n} \\
\bar{F}^{s}_{x_1,x_2;k_y;i\omega_n} & \bar{G}^{s}_{x_1,x_2;k_y;i\omega_n}
\end{array} \right), \\
\hat{\mathcal{G}}_{ij}(k_y;i\omega_n)&=\left(
\begin{array}{cc}
G^n_{i,j;k_y;i\omega_n} & F^n_{i,j;k_y;i\omega_n} \\
\bar{F}^n_{i,j;k_y;i\omega_n} & \bar{G}^n_{i,j;k_y;i\omega_n}
\end{array} \right),
\end{aligned}
\label{eq:G_nambu}
\end{equation}
where, we have Fourier-transformed from imaginary time $\tau$ to Matsubara frequency $i\omega_n$ and we note that for spin-independent normal states and spin-singlet superconductors: $\bar{G}^{s}_{x_1,x_2;k_y;i\omega_n}=-(G^{s}_{x_1,x_2;-k_y;i\omega_n})^*$, $\bar{F}^{s}_{x_1,x_2;k_y;i\omega_n}=(F^{s}_{x_1,x_2;-k_y;i\omega_n})^*$, $\bar{G}^{n}_{i,j;k_y;i\omega_n}=-\left(G^n_{i,j;-k_y;i\omega_n}\right)^*$, and $\bar{F}^{n}_{i,j;k_y;i\omega_n}=\left(F^n_{i,j;-k_y;i\omega_n}\right)^*$. 

In the absence of tunneling between the nanowires and the superconductor, i.e.~$t_L=t_R=0$, it is straightforward to show that the Green's functions in Eqs.~(\ref{eq:G_nambu}) are given by:
\begin{equation}
\begin{aligned}
\hat{\mathcal{G}}^{(0)}_{s}(x,k_y;i\omega_n)&=\left[i\omega_n\hat{\tau}_0+\Delta\hat{\tau}_1\right]g_0(x,k_y,i\omega_n) \\
&+g_3(x,k_y,i\omega_n)\hat{\tau}_3, \\
\hat{\mathcal{G}}^{(0)}_{ij}(k_y;i\omega_n)&=-\delta_{ij}\frac{i\omega_n\hat{\tau}_0+\xi_{i,k_y}\hat{\tau}_3}{\omega_n^2+\xi_{i,k_y}^2},
\end{aligned}
\label{eq:g0}
\end{equation}
where the coefficients, $g_0$ and $g_3$, are given by Eqs (\ref{eq:g0g3f_matsubara_appendix}) in the Appendix, and $\hat{\tau}_0$ and $\hat{\tau}_i$ are the 2$\times$2 identity and Pauli matrices in particle-hole space, respectively. While the exact forms of $g_0$ and $g_3$ are less important, we note that both ere even in $x$, $k_y$, and $\omega_n$: $g_0(x,k_y,i\omega_n)=g_0(-x,-k_y,-i\omega_n)$ and $g_3(x,k_y,i\omega_n)=g_3(-x,-k_y,-i\omega_n)$. 

At finite tunneling, $t_L,t_R\neq 0$, the Green's functions in Eqs.~(\ref{eq:G_nambu}) satisfy the following Dyson equations:
\begin{equation} 
\begin{aligned}
\hat{\mathcal{G}}_{s}(x_1,x_2)&=\hat{\mathcal{G}}^{(0)}_{s}(x_1-x_2)+\hat{\mathcal{G}}^{(0)}_{s}(x_1-x_L) \hat{\Sigma}^{L}_s\hat{\mathcal{G}}_{s}(x_L,x_2) \\
&+\hat{\mathcal{G}}^{(0)}_{s}(x_1-x_R) \hat{\Sigma}^{R}_s\hat{\mathcal{G}}_{s}(x_R,x_2), \\
\hat{\mathcal{G}}_{ij}&=\hat{\mathcal{G}}^{(0)}_{ij} + \hat{\mathcal{G}}^{(0)}_{ii}\hat{\Sigma}^{n}_{iL}\hat{\mathcal{G}}_{Lj} + \hat{\mathcal{G}}^{(0)}_{ii}\hat{\Sigma}^{n}_{iR}\hat{\mathcal{G}}_{Rj}, \\
\end{aligned}
\label{eq:dyson}
\end{equation}
with the self-energies defined as:
\begin{equation}
\begin{aligned}
\hat{\Sigma}^{i}_s&=t_i^2\hat{\tau}_3\hat{\mathcal{G}}^{(0)}_{ii}\hat{\tau}_3, \\
\hat{\Sigma}^{n}_{ij}&=t_it_j\hat{\tau}_3\hat{\mathcal{G}}^{(0)}_{s}(x_i-x_j)\hat{\tau}_3,
\end{aligned}
\label{eq:self}
\end{equation}
where we have omitted the explicit dependence on $k_y$ and $i\omega_n$, since both of these quantities are conserved.

\section{Odd-frequency Pairing}
\label{sec:odd}

By iterating Eq.~(\ref{eq:dyson}) we can compute the Green's functions, $\hat{\mathcal{G}}_{s}$ and $\hat{\mathcal{G}}_{ij}$, in terms of the bare Green's functions given in Eq.~(\ref{eq:g0}) to arbitrary order in powers of the tunneling parameters $t_i$. In the limit of weak coupling between the nanowires and the superconducting substrate, the physics is dominated by the leading order terms in $t_i$ and we have:
\begin{equation} 
\begin{aligned}
\hat{\mathcal{G}}_{s}(x_1,x_2)&\approx\hat{\mathcal{G}}^{(0)}_{s}(x_1-x_2)+\delta\hat{\mathcal{G}}^{(1)}_{s}(x_1,x_2), \\
\hat{\mathcal{G}}_{ij}&\approx\hat{\mathcal{G}}^{(0)}_{ij} + \delta\hat{\mathcal{G}}^{(1)}_{ij},
\end{aligned}
\label{eq:series_1}
\end{equation}
where 
\begin{equation} 
\begin{aligned}
\delta\hat{\mathcal{G}}^{(1)}_{s}(x_1,x_2)&=\sum_{i=L,R}\hat{\mathcal{G}}^{(0)}_{s}(x_1-x_i) \hat{\Sigma}^{i}_s\hat{\mathcal{G}}^{(0)}_{s}(x_i-x_2), \\
\delta\hat{\mathcal{G}}^{(1)}_{ij}&= \hat{\mathcal{G}}^{(0)}_{ii}\hat{\Sigma}^{n}_{ij}\hat{\mathcal{G}}^{(0)}_{jj}.
\end{aligned}
\label{eq:delta_1}
\end{equation}
By inserting the expressions from Eqs.~(\ref{eq:g0}) into Eqs.~(\ref{eq:delta_1}) we can explicitly calculate the leading order corrections to the Green's functions in both the nanowires and the superconducting substrate. Without loss of generality, we assume, for concreteness, that $x_L=-d/2$, $x_R=d/2$ in Eq.~(\ref{eq:delta_1}). To study the superconducting pairing, we only need to focus on the anomalous parts of the Green's functions in Eqs.~(\ref{eq:delta_1}) and find:
\begin{widetext}
\begin{equation}
\begin{aligned}
\delta F^{(1)}_{s}&(x_1,x_2)=\frac{4m_s^4\Delta}{\Omega_n^2k_0^2}\left[i\omega_n \frac{\Omega_n}{2m_s} A_{x_1,x_2}(i\omega_n) + \omega_n^2B_{x_1,x_2}(i\omega_n) +\frac{\Omega_n}{2m_s} C_{x_1,x_2}(i\omega_n) \right], \\
\delta F^{(1)}_{ij}&=t_it_j g_{0}(x_i-x_j,k_y,i\omega_n) \Delta  \frac{\left(\omega_n^2 + \xi_{i,k_y}\xi_{j,k_y}\right)-i\omega_n\left( \xi_{i,k_y}-\xi_{j,k_y}\right)}{\left(\omega_n^2+\xi_{i,k_y}^2\right)\left(\omega_n^2+\xi_{j,k_y}^2\right)},
\end{aligned}
\label{eq:dF1}
\end{equation}
\end{widetext}
where we have defined $\Omega_n=2m_s\sqrt{\omega_n^2+\Delta^2}$ and $k_0=\left(\alpha^2+\Omega_n^2\right)^{1/4}$, with $\alpha=2m_s\mu_s-k_y^2$, and the functions $A_{x_1,x_2}(i\omega_n)$, $B_{x_1,x_2}(i\omega_n)$, and $C_{x_1,x_2}(i\omega_n)$ are given by Eqs (\ref{eq:A})-(\ref{eq:C}) in the Appendix. We will consider their behavior in more detail below but, for now, we only note they are all even functions of $i\omega_n$ and thus that the presence of the nanowires modifies the pair amplitudes within the substrate, inducing both novel even-$\omega$ pair amplitudes, proportional to $B$ and $C$, and odd-$\omega$ pair amplitudes, proportional to $A$. Additionally, we notice that the proximity-induced pairing within the nanowires $\delta F_{ij}^{(1)}$, possesses both an even-$\omega$ term and an odd-$\omega$ term. Furthermore, while the even-$\omega$ term is non-zero in both the intrawire and interwire channels, the odd-$\omega$ terms belong strictly to the interwire channel. By permuting the wire index, it is easy to see that the odd-$\omega$ pair amplitude in the nanowires is also odd in the wire index, consistent with the constraints imposed by Fermi-Dirac statistics.\cite{triolaprb2016}  

\subsection{Odd-$\omega$ pairing in the nanowires} 
Having demonstrated in Eq.~(\ref{eq:dF1}) that multiple odd-$\omega$ pair amplitudes can be induced in a double wire system, we now study the nature of these odd-$\omega$ correlations in more depth, starting with the proximity-induced interwire pairing. In this subsection we will continue to present results in $k_y$-space because translation invariance is preserved in the $y$-direction and, importantly, the resulting expressions are easier to understand when resolved in $k_y$.

From Eq.~(\ref{eq:dF1}) it is clear that, in general, the interwire pair amplitude possesses both even-$\omega$ and odd-$\omega$ terms. Moreover, since essentially all of the complications arising from the coupling to the superconducting substrate take the form of a multiplicative prefactor, it is quite easy to obtain the ratio of the odd-$\omega$ pairing to the even-$\omega$ pairing, given by:
\begin{equation}
\delta F_{o:e}(k_y;\omega)=\frac{(\omega+i\eta)\left( \xi_{L,k_y}-\xi_{R,k_y}\right)}{(\omega+i\eta)^2 - \xi_{L,k_y}\xi_{R,k_y}},
\label{eq:f_odd/even_real}
\end{equation}
where we have performed the analytic continuation to real frequency, $i\omega_n\rightarrow \omega+i\eta$, to make contact with the physical spectrum of the system. This simple ratio allows us to determine the precise conditions for which we expect the odd-$\omega$ pair amplitudes to dominate over the even-$\omega$ amplitudes. Moreover, due to the properties of the Fourier transform, we note that, by evaluating this expression at $k_y=0$, we obtain the ratio of the total, i.e.~integrated, odd-$\omega$ pairing in real space to the total even-$\omega$ pairing in real space: $\delta F_{o:e}(k_y=0;\omega)=\int_{-\infty}^\infty F_{odd}(y;\omega)dy/\int_{-\infty}^\infty F_{even}(y;\omega)dy$.
  
\begin{figure}
 \begin{center}
  \centering
        \includegraphics[width=0.5\textwidth]{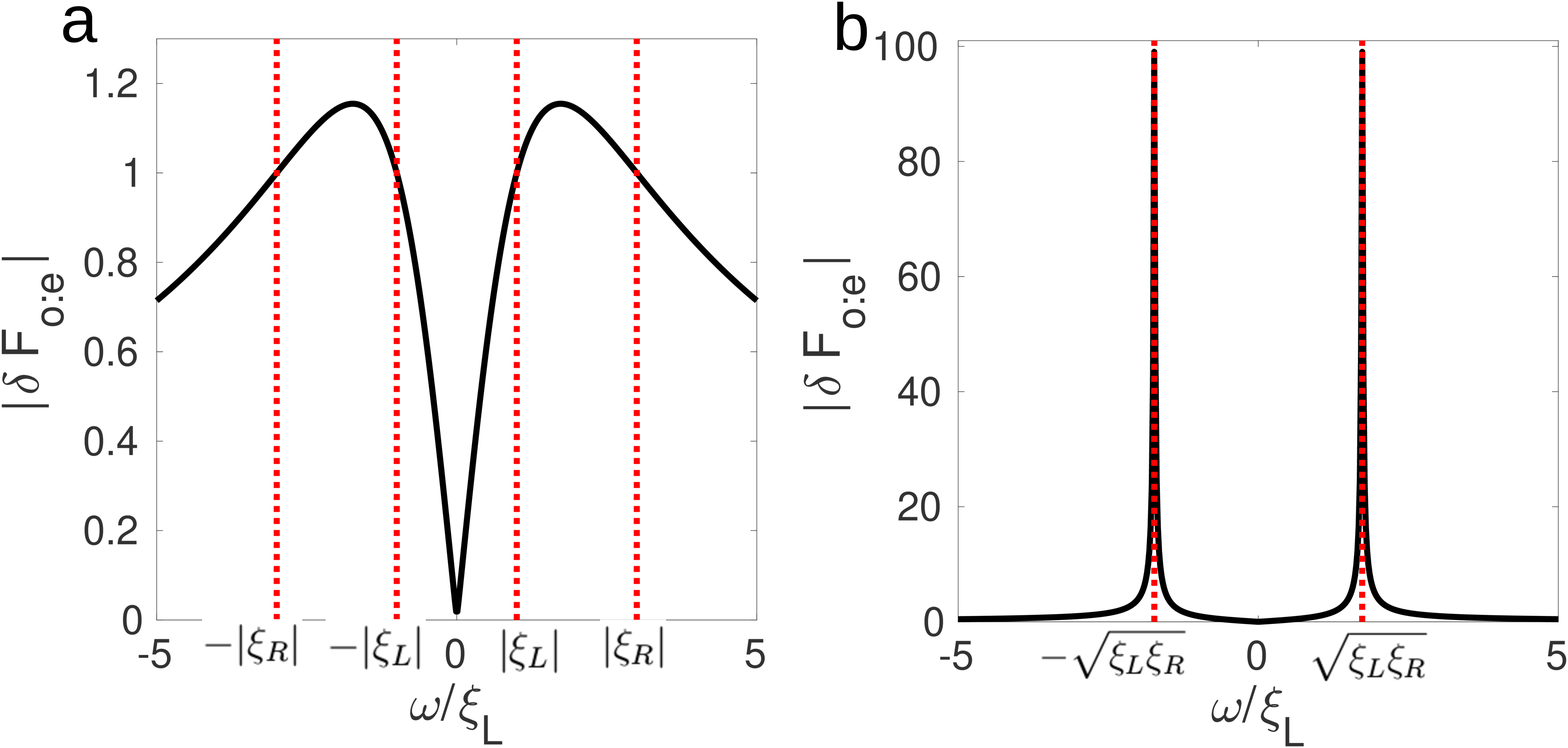}
  \caption{Absolute magnitude of the ratio of odd-$\omega$ to even-$\omega$ pair amplitudes in the interwire channel $|\delta F_{o:e}|$, computed using Eq.~(\ref{eq:f_odd/even_real}), with $k_y$ fixed and units chosen such that $\xi_L=1$.
  (a) $|\delta F_{o:e}(\omega)|$ for $\xi_R=-3\xi_L$, with vertical dotted lines indicating frequencies for which the ratio equals unity: $\omega=\{\pm\xi_{L,k_y}, \pm\xi_{R,k_y}\}$. (b) $|\delta F_{o:e}(\omega)|$ for $\xi_R=3\xi_L$, with vertical dotted lines indicating frequencies with strictly odd-$\omega$ pairing: $\omega=\pm\sqrt{\xi_{L,k_y}\xi_{R,k_y}}$. 
  }
  \label{fig:ratio}
 \end{center}
\end{figure}

From Eq.~(\ref{eq:f_odd/even_real}) it is clear that the odd-$\omega$ amplitude will be non-zero as long as $\xi_{L,k_y}\neq\xi_{R,k_y}$, and that the two pair symmetries will be equal in magnitude at the frequencies $\omega=\{\pm\xi_{L,k_y}, \pm\xi_{R,k_y}\}$, see Fig.~\ref{fig:ratio}(a) for an example. Furthermore, it is clear that the even-$\omega$ pair amplitudes vanish exactly at $\omega=\sqrt{\xi_{L,k_y}\xi_{R,k_y}}$. Therefore, so long as $\xi_{L,k_y}\neq\xi_{R,k_y}$, and $\xi_{L,k_y}$ and $\xi_{R,k_y}$ possess the same sign, the interwire pairing will be strictly odd-$\omega$ at $\omega=\pm\sqrt{\xi_{L,k_y}\xi_{R,k_y}}$, as illustrated in Fig.~\ref{fig:ratio}(b). This pure odd-$\omega$ interwire pairing criteria can be engineered by adjusting the chemical potentials within the two nanowires, for example by electrostatic gating.   

Having discussed the relative size of the even-$\omega$ and odd-$\omega$ interwire pair amplitudes, we now turn our attention to the overall magnitude of the interwire pairing. First, comparing the above criteria for an odd-$\omega$-dominated interwire channel to the expressions in Eq.~(\ref{eq:dF1}), we see that the frequencies for which $|\delta F_{o:e}|>1$ align precisely with the poles in the denominator of the total interwire pair amplitude. Therefore, the denominator should not have a deleterious effect on the odd-$\omega$ pairing. 
Then, neglecting the denominator in Eq.~(\ref{eq:dF1}), we see that the magnitude of the interwire pairing is determined by only three factors: $t_i$, $\Delta$, and $g_{0}$. The hopping amplitudes $t_i$ depend sensitively on the microscopic model of the nanowire-superconductor interface. A precise determination of their values is clearly beyond the scope of this work and, for our purposes, these parameters are simply constants characterizing the interface. Further, the gap of the superconducting substrate $\Delta$ may be adjusted by choosing different substrates, and is therefore an external parameter. It is thus the function $g_{0}$ that carries all the relevant dependences for the interwire pairing, such as information about the kinetic energy of the substrate, as well as the distance between the two nanowires.

Since we are primarily interested in interwire pairing, we focus on evaluating the function $g_0(x_i-x_j)$ for $i\neq j$, and since $g_0(x)=g_0(-x)$ we arrive at:
\begin{equation}
g_{0}(d)=-\frac{2 m_s^2e^{-|d|k_0\sin\tfrac{\phi}{2}}}{k_0\Omega_n}\cos\left(|d|k_0\cos\tfrac{\phi}{2}-\tfrac{\phi}{2} \right) ,
\label{eq:f_LR}
\end{equation}  
where we have suppressed the dependence on $k_y$ and $i\omega_n$ on the left-hand side, and we have defined $\phi=\arctan(\Omega_n/\alpha)$ using $\Omega_n$ and $\alpha$ as given below Eq. (\ref{eq:dF1}). From Eq.~(\ref{eq:f_LR}) we readily see that for large nanowire separations, $d>>1/k_0$, the magnitude of the interwire pair amplitudes go as $\sim e^{-|d|k_0\sin{\tfrac{\phi}{2}}}$. Furthermore, setting $k_y=0$ to obtain the average of this quantity over the $y$-axis, and assuming $\mu_s>>\sqrt{\omega_n^2+\Delta^2}$, we find that this exponential decay factor becomes $e^{-|d|\tfrac{\Delta}{v_F}\sqrt{1+\omega_n^2/\Delta^2}}$, where $v_F=\sqrt{2\mu_s/m_s}$, which is consistent with the expectation that the pair correlations should be suppressed when $d$ exceeds the superconducting coherence length, $\xi\sim v_F/\Delta$.

\subsection{Odd-$\omega$ pairing in the superconducting substrate}
\label{subsec:oddwsub}
We next turn our attention to the pair symmetry of the superconducting substrate. The corrections to the anomalous Green's function in the substrate due to the presence of the nanowires are given by Eq.~(\ref{eq:dF1}), written in terms of the coefficients $A$, $B$, and $C$, which are given in the appendix, Eqs.~(\ref{eq:A})-(\ref{eq:C}). 

By inspecting the functions in Eqs.~(\ref{eq:A})-(\ref{eq:C}), we immediately see that all three are even in Matsubara frequency $i\omega_n$ and $k_y$, since they only depend on these variables through $\omega_n^2$ and $k_y^2$, respectively. Therefore, we find that the only odd-$\omega$ term in the anomalous Green's function Eq.~(\ref{eq:dF1}) is the term proportional to $A$. Furthermore, from Eqs.~(\ref{eq:A})-(\ref{eq:C}), we observe that the spatial parities of these coefficients under the exchange of the $x$-coordinates are: $A_{x_1,x_2}=-A_{x_2,x_1}$, $B_{x_1,x_2}=B_{x_2,x_1}$, and $C_{x_1,x_2}=C_{x_2,x_1}$. Since, the odd-$\omega$ amplitude is proportional to $A$ and the even-$\omega$ amplitudes are proportional to $B$ and $C$, we see that these symmetries are fully consistent with the constraints imposed by Fermi-Dirac statistics.    

Another notable feature of the expressions for $A$, $B$, and $C$, in Eqs.~(\ref{eq:A})-(\ref{eq:C}), is that, while the coordinate dependence is somewhat complicated in general, we can see that when $x_1$ and $x_2$ are sufficiently far from both nanowires, i.e.~$|x-x_L|,|x-x_R|>>v_F/\sqrt{\omega_n^2+\Delta^2}$, all corrections are exponentially suppressed when averaged over the length of the nanowires. For low frequencies, this length scale is proportional to the coherence length of the bare substrate. Interestingly, there is no preferential suppression of the odd-$\omega$ terms coming from the exponential factors; even-$\omega$ and odd-$\omega$ amplitudes get comparably suppressed, similar to the proximity-induced pairing in the nanowires discussed in the previous subsection.

\section{Higher-order pair symmetry conversion}
\label{sec:high}

In the previous section we demonstrated that odd-$\omega$ pair amplitudes are induced in both the superconducting substrate and the interwire channel of the nanowires. These results were obtained using a perturbative expansion in the hopping amplitudes $t_i$ between the nanowires and the substrate. The benefit of such a calculation is that it is relatively simple and the symmetries of the pair amplitudes are made manifest. However, such an analysis is limited to small values of $t_i$, as it ignores higher-order terms in the expansion. In this section we instead solve the problem exactly using a $T$-matrix approach and thus incorporate the effect of all higher-order tunneling processes on the pair amplitudes. We compare our exact results to the perturbative ones in the previous section, and most importantly, demonstrate novel features of the pair symmetry which only emerge at higher orders. In particular, we focus on the way in which higher-order processes can allow the odd-$\omega$ odd-parity pairing to play a role in the local properties of the system. As we will show this effect can be very important for the superconducting pairing in the substrate. However, since the interwire pairing, by its very nature, requires coupling to both wires, we do not expect higher-order terms to allow the interwire odd-$\omega$ amplitudes to contribute significantly to any local observables. Such an effect might be more relevant in a setup possessing three or more nanowires, but such an analysis is clearly beyond the scope of our current work. Therefore, for the remainder of this work we limit ourselves to pair amplitudes within the substrate, which we already discussed within the perturbative weak coupling limit in subsection \ref{subsec:oddwsub}.

To perform our analysis, we return to the Dyson equation describing the exact Green's functions of the superconducting substrate, Eq.~(\ref{eq:dyson}). By iterating this equation we find that above order $t^2$, cross-terms begin to emerge which involve a propagation between the nanowires of the form, $\hat{\mathcal{G}}^{(0)}_{s}(x_1-x_L) \hat{\Sigma}^{L}_s\hat{\mathcal{G}}^{(0)}_{s}(x_L-x_R)\hat{\Sigma}^{R}_s \hat{\mathcal{G}}^{(0)}_{s}(x_R-x_2)$. At higher orders more of these terms emerge, thus significantly complicating an evaluation using a $T$-matrix. To alleviate this problem, we start by neglecting the right nanowire and exactly solve the problem of the superconducting substrate coupled to the left nanowire (L+SC) only. We then turn our attention to the combined L+SC system in the presence of the right nanowire and solve the problem (L+SC+R) exactly. In this way we account for all cross-terms while still being able to proceed analytically. 

Since we are dealing with Green's functions whose arguments are a mixture of momentum, $k_y$, and positions, $x_1$,$x_2$, the position of the poles for these functions depend on the complex-valued frequency and the momentum, $k_y$. To keep track of both the Matsubara and retarded Green's functions we derive all expressions in this section for Green's functions with a generic complex frequency, $z$. In this way, all results apply equally well to Matsubara, retarded, and advanced Green's functions.

\subsection{Left nanowire + superconductor}
\label{sec:L+SC}
In the presence of only the left nanowire it is straightforward to show that Eq.~(\ref{eq:dyson}) can be written as:
\begin{equation} 
\hat{\mathcal{G}}^{(L)}_{s}(x_1,x_2)=\hat{\mathcal{G}}^{(0)}_{s}(x_1-x_2)+\hat{\mathcal{G}}^{(0)}_{s}(x_1+\tfrac{d}{2}) \hat{T}_{L}\hat{\mathcal{G}}^{(0)}_{s}(x_2+\tfrac{d}{2}) 
\label{eq:G_L}
\end{equation}
where the $T$-matrix is defined as
\begin{equation}
\hat{T}_{L}=\left[(\hat{\Sigma}^{L}_s)^{-1}- \hat{\mathcal{G}}^{(0)}_{s}(0)\right]^{-1}.
\label{eq:T_L}
\end{equation}
The bare Green's function of the substrate appearing in Eqs.~(\ref{eq:G_L}) and (\ref{eq:T_L}), $\hat{\mathcal{G}}^{(0)}_{s}(x)$, is a function of position $x$, momentum $k_y$, and frequency $z$ with the general structure given by
\begin{equation}
\hat{\mathcal{G}}^{(0)}_{s}(x)=\left[z\hat{\tau}_0+\Delta\hat{\tau}_1\right]g_0(x)+g_3(x)\hat{\tau}_3,
\label{eq:schematic_g0}
\end{equation}  
where the coefficients $g_0$ and $g_3$ are complicated functions of $|x|$, $k_y^2$, and $z^2$, given in Appendix \ref{app:g0}, see Eqs.~(\ref{eq:g0g3f_matsubara_appendix}) for $z=i\omega$, and Eqs.~(\ref{eq:g0g3f_appendix}) for $z=\omega+i\eta$. However, for compactness we suppress the $k_y$ and $z$ arguments as they will not change throughout this derivation. Importantly, the functional dependences imply that both $g_0$ and $g_3$ are even under the transformations: $x\rightarrow -x$, $k_y\rightarrow-k_y$, and $z\rightarrow -z$. Next, using the general form in Eq.~(\ref{eq:schematic_g0}) together with the definition of $\hat{\Sigma}^{L}_s$ in Eq.~(\ref{eq:self}), it is straightforward to show that the $T$-matrix in Eq.~(\ref{eq:T_L}) takes the form:
\begin{equation}
\hat{T}_{L}=zT_0\hat{\tau}_0+T_3\hat{\tau}_3 +T_1\hat{\tau}_1,
\label{eq:TL_explicit}
\end{equation}
with the coefficients given in Appendix \ref{app:tmatrix}. While the expressions in Eq.~(\ref{eq:T013}) are somewhat complicated, we notice that these functions inherit the symmetries of $g_0$ and $g_3$, and are, hence, even under the transformations: $k_y\rightarrow-k_y$ and $z\rightarrow -z$. 

Inserting Eq.~(\ref{eq:TL_explicit}) into Eq.~(\ref{eq:G_L}), we arrive at the Green's function of the L+SC system to infinite order in the tunneling $t_L$:
\begin{widetext}
\begin{equation}
\hat{\mathcal{G}}^{(L)}_{s}(x_1,x_2)=zg^{(L)}_0(x_1,x_2)\hat{\tau}_0 + g^{(L)}_3(x_1,x_2)\hat{\tau}_3 + f^{(L)}_1(x_1,x_2)\hat{\tau}_1 + izf^{(L)}_2(x_1,x_2)\hat{\tau}_2, 
\label{eq:G_L_explicit}
\end{equation}
\end{widetext}
where we have defined the coefficients $g^{(L)}_0$, $g^{(L)}_3$, $f^{(L)}_1$, and $f^{(L)}_2$, given in Eq.~(\ref{eq:GL0123}). By inspecting these terms, it is clear that all of the coefficients $g^{(L)}_0$, $g^{(L)}_3$, $f^{(L)}_1$, and $f^{(L)}_2$ are invariant under the transformations: $z\rightarrow -z$, $k_y\rightarrow-k_y$. However, by permuting the coordinate indices, $x_1\leftrightarrow x_2$, we find that $g^{(L)}_0(x_1,x_2)=g^{(L)}_0(x_2,x_1)$, $g^{(L)}_3(x_1,x_2)=g^{(L)}_3(x_2,x_1)$, and $f^{(L)}_1(x_1,x_2)=f^{(L)}_1(x_2,x_1)$, while $f^{(L)}_2(x_1,x_2)=-f^{(L)}_2(x_2,x_1)$. 

To study the pair symmetries, we focus on the anomalous part of the Green's function given by Eq.~(\ref{eq:G_L_explicit}) and find:
\begin{equation}
F^{(L)}_s(x_1,x_2)=f^{(L)}_1(x_1,x_2)+zf^{(L)}_2(x_1,x_2).
\label{eq:FL}
\end{equation}
This equation shows how the presence of the nanowire changes the pair amplitudes in the superconducting substrate, both altering the even-$\omega$ channel, given by $f^{(L)}_1$, and inducing an odd-$\omega$ channel, proportional to $f^{(L)}_2$. Noting the symmetries of these two functions, we see that the even-$\omega$ channel is even under the permutation of the coordinates, $x_1\leftrightarrow x_2$, while the odd-$\omega$ channel is odd under this permutation, consistent with the constraints imposed by Fermi-Dirac statistics as both amplitudes are necessarily spin-singlet states.

By inserting the expressions for $g_0$ and $g_3$, from Appendix \ref{app:g0}, we readily obtain exact expressions for the even-$\omega$, $F^{(L)}_{even}=f^{(L)}_1(x_1,x_2)$, and odd-$\omega$, $F^{(L)}_{odd}=zf^{(L)}_2(x_1,x_2)$, pair amplitudes existing within the superconducting substrate coupled to a single nanowire located at $x=-d/2$. While the expressions for $F^{(L)}_{even}$ are quite cumbersome and not very enlightening, we can gain analytical insight by inspecting the odd-$\omega$ pair amplitude, given by: 
\begin{widetext}
\begin{equation}
F^{(L)}_{odd}(x_1,x_2)=zf^{(L)}_2(x_1,x_2)=\frac{z\Delta t_L^2\left[g_3(x_1+\tfrac{d}{2})g_0(x_2+\tfrac{d}{2})-g_0(x_1+\tfrac{d}{2})g_3(x_2+\tfrac{d}{2}) \right]}{\left[z^2-\xi_{L,k_y}^2\right]-2t_L^2\left[z^2g_0(0)+\xi_{L,k_y}g_3(0)\right]-t_L^4\left[ (\Delta^2-z^2)g_0^2(0)+g_3^2(0) \right]}. 
\label{eq:FL_odd}
\end{equation}
\end{widetext}
By exchanging the coordinates $x_1$ and $x_2$ we can readily verify that $F^{(L)}_{odd}(x_1,x_2)=-F^{(L)}_{odd}(x_2,x_1)$, as mentioned above. Additionally, we note that $F^{(L)}_{odd}(x_1,x_2)$ is proportional to $\Delta g_0$, which is the anomalous Green's function of the bare substrate. Therefore, we conclude that the odd-$\omega$ pair amplitude is heavily peaked at $\omega=\Delta$ and decays for $\omega>\Delta$. Furthermore, from the denominator we infer that $F^{(L)}_{odd}(x_1,x_2)$ obtains its largest contribution when the frequencies match the energy levels in the nanowire, $z=\xi_{L,k_y}$. Combining these two insights we determine that the odd-$\omega$ amplitude will be maximized when $\mu_L<\Delta$ and decrease for $\mu_L>>\Delta$.

Further insight can be gained in the limit of weak coupling between the nanowire and the substrate, $t_L/\mu_s<<1$. Then, assuming realistically that $\Delta/\mu_s<<1$ and focusing on frequencies $z=\omega +i\eta$, we evaluate Eq.~(\ref{eq:FL_odd}) locally along the $y$-axis:
\begin{widetext}
\begin{equation}
F^{(L)}_{odd}(x_1,x_2,y=0;\omega)\approx \frac{i\text{sgn}(\eta)t_L^2 \sqrt{2m_L} \Delta \sin\left[\left(|x_1+\tfrac{d}{2}|-|x_2+\tfrac{d}{2}|\right)k_F\right]\left(\sqrt{\mu_L+(\omega+i\eta)}+\sqrt{\mu_L-(\omega+i\eta)} \right)}{4v_F^2\sqrt{\Delta^2-(\omega+i\eta)^2}\sqrt{\mu_L^2-(\omega+i\eta)^2}\exp\left[\frac{|x_1+\tfrac{d}{2}|+|x_2+\tfrac{d}{2}|}{v_F}\sqrt{\Delta^2-(\omega+i\eta)^2}\right]}, 
\label{eq:FL_odd_approx}
\end{equation}
\end{widetext}
where $k_F=\sqrt{2m_s\mu_s}$ and $v_F=k_F/m_s$. We can see that this amplitude is odd in $z=\omega+i\eta$ since it is proportional to $\text{sgn}(\eta)$. Also, we see that it possesses most of its weight near $\omega=\pm\Delta$ and $\omega=\pm\mu_L$. Centering the coordinates on the nanowire, we see that the absolute magnitude of the odd-$\omega$ pair amplitude possesses local maxima at $x=|x_1|-|x_2|=(2n+1)\pi/(2 k_F)$, where $n\in\mathbb{Z}$. Moreover, this amplitude is fairly long-ranged at frequencies close to the gap, and decays as $\sim \exp\left[-\Delta\left(\tfrac{|x_1|+|x_2|}{v_F}\right)\right]$ for frequencies below the gap.

\begin{figure}
 \begin{center}
  \centering
        \includegraphics[width=0.5\textwidth]{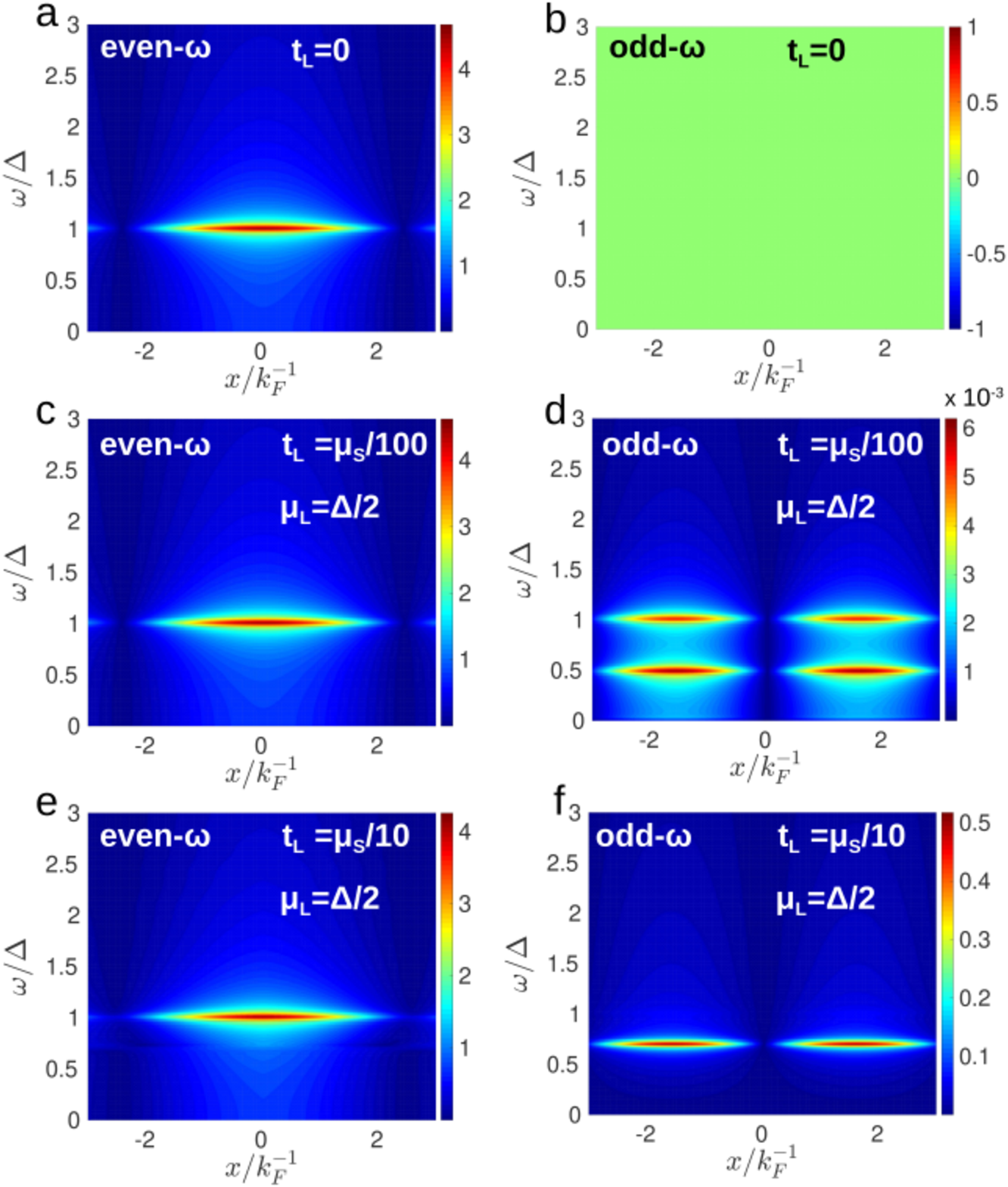}
  \caption{Magnitude of even-$\omega$ (left column) and odd-$\omega$ (right column) terms in the retarded pair amplitude of the superconducting substrate, $|F^{(L)}_s(x-\tfrac{d}{2},-\tfrac{d}{2},z=\omega+i\eta)|$, plotted as a function of the distance from the nanowire $x$, and the frequency above the Fermi level $\omega$. Amplitudes are computed by numerically integrating the exact expression in Eq.~(\ref{eq:FL}) over $k_y$. We set the parameters of the substrate so that $\Delta=\mu_s/100$, and we report distances in $k_F^{-1}$, where $k_F=\sqrt{2m_s\mu_s}$. For all plots $m_L=m_s$ and $\mu_L=\Delta/2$, while the three rows represent different tunneling parameter choices: $t_L=0$ (a,b), $t_L=\mu_s/100$ (c,d), and $t_L=\mu_s/10$(e,f).}
  \label{fig:L+SC_xw}
 \end{center}
\end{figure}

To confirm these conclusions about the odd-$\omega$ pair amplitudes, and gain further insight into the behavior of both the even-$\omega$ and odd-$\omega$ pairing in this system, we plot in Fig.~\ref{fig:L+SC_xw} the absolute magnitudes of both the even-$\omega$ and odd-$\omega$ pair amplitudes given as functions of distance from the nanowire position, $|F^{(L)}_s(x-\tfrac{d}{2},-\tfrac{d}{2};z=\omega+i\eta)|$. To make contact with the real spectrum, we use the retarded Green's functions in Appendix \ref{app:g0}, and we now eliminate the momentum dependence by integrating all expressions over $k_y$. 

In Figs.~\ref{fig:L+SC_xw}(a, b) we see that, in the absence of tunneling between the nanowire and the substrate, the odd-$\omega$ pair amplitude is always zero, while the even-$\omega$ pair amplitude is heavily peaked around the point ($x=0$, $\omega=\Delta$) i.e.~locally and at the energy of the bare coherence peaks. This is completely consistent with our expectations of the bare Green's functions for the substrate. In Figs.~\ref{fig:L+SC_xw}(c, d) we set the tunneling between the nanowire and the substrate to a value of $t_L=\mu_s/100$ and find that both the even-$\omega$ and odd-$\omega$ amplitudes are now non-zero. The even-$\omega$ amplitudes remain peaked around ($x=0$, $\omega=\Delta$), while the odd-$\omega$ amplitudes possess peaks at $x=\pm\pi/2k_F^{-1}$ and near $\omega=\pm\mu_L,\pm\Delta$, in precise agreement with the analytic results above. Finally in Figs.~\ref{fig:L+SC_xw} (e,f), we increase the tunneling by a a factor of 10 to $t_L=\mu_s/10$ and see that the odd-$\omega$ amplitude has increased by a factor of 100, consistent with its leading-order $t_L^2$ dependence, while the overall behavior of the even-$\omega$ is qualitatively unchanged. Additionally, we notice that for this larger value of $t_L$ the peaks in the odd-$\omega$ amplitude still occur at $x=\pm\pi/2k_F^{-1}$ but they are now sharply peaked around $\omega\approx 0.7\Delta$, lying roughly midway between the peaks predicted from the weak-tunneling results. 

\begin{figure}
 \begin{center}
  \centering
        \includegraphics[width=0.45\textwidth]{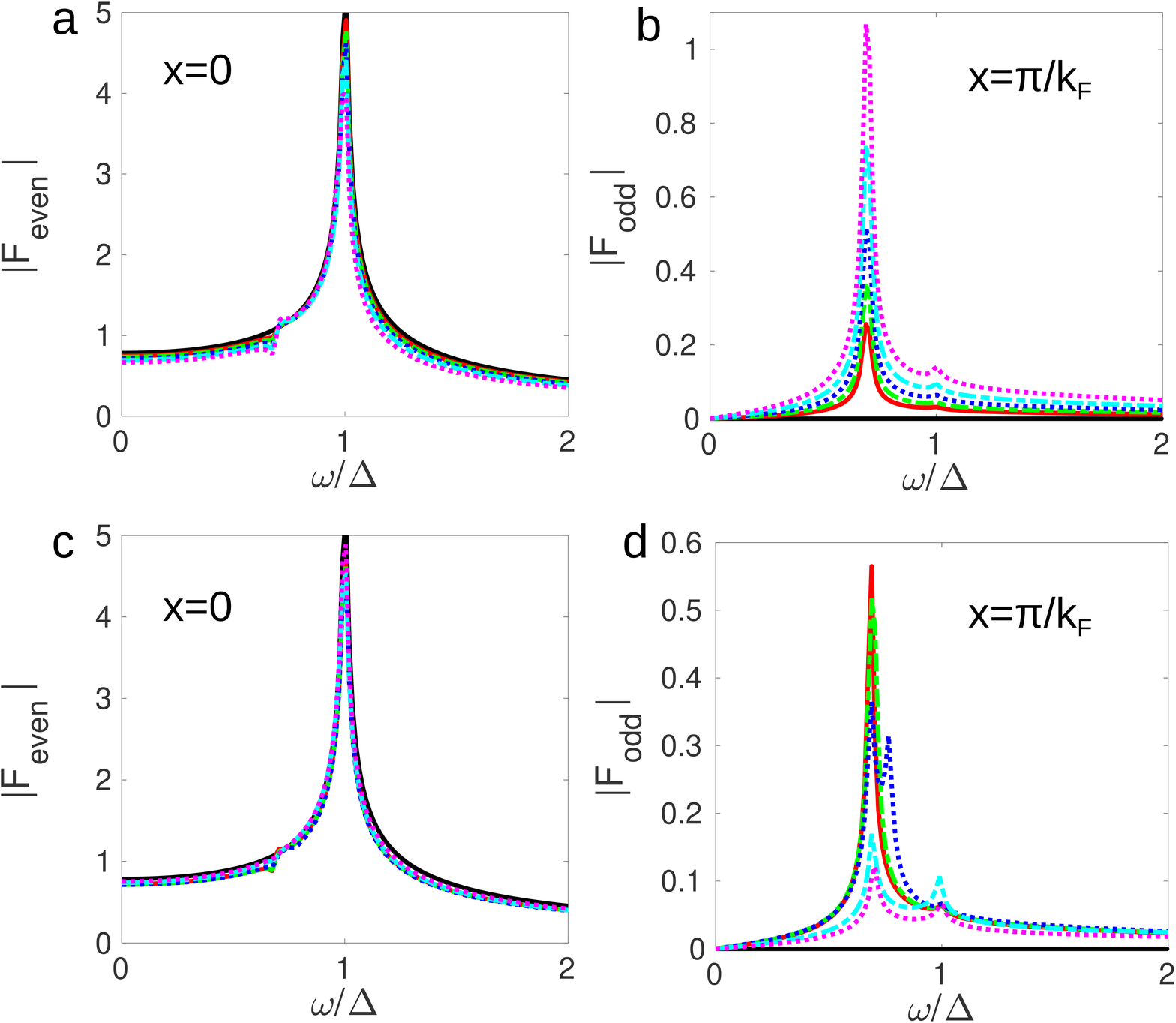}
  \caption{Magnitude of even-$\omega$ (left column) and odd-$\omega$ (right column) terms in the retarded pair amplitude of the superconducting substrate, $|F^{(L)}_s(x-\tfrac{d}{2},-\tfrac{d}{2},z=\omega+i\eta)|$, plotted as a function of frequency, $\omega$ and evaluated at the positions where the pair amplitude peaks, $x=0$ and $x=\pi/(2k_F)$, respectively, where $k_F=\sqrt{2m_s\mu_s}$. The amplitudes are computed by numerically integrating the exact expression in Eq.~(\ref{eq:FL}) over $k_y$. 
We set the parameters of the substrate so that $\Delta=\mu_s/100$, and we have fixed $t_L=\mu_s/10$ to obtain a large odd-$\omega$ component. 
(a,b): Fixed $\mu_L=\Delta/2$ for different masses, $m_L=m_S/4$ (solid/red), $m_L=m_S/2$ (dashed-dotted/green), $m_L=m_S$ (dashed-dotted/cyan), $m_L=2m_S$ (dotted/blue), and $m_L=4m_S$ (dotted/magenta). 
(c,d): Fixed $m_L=m_S$ for different chemical potentials, $\mu_L=\Delta/4$ (solid/red), $\mu_L=\Delta/2$ (dashed/green), $\mu_L=\Delta$ (dashed-dotted/cyan), $\mu_L=5\Delta$ (dashed-dotted/blue),  and $\mu_L=10\Delta$ (dotted/magenta). In each panel we also plot the results for $t_L=0$ (solid/black). }
  \label{fig:L+SC_x}
 \end{center}
\end{figure}

In Fig.~\ref{fig:L+SC_x} we further examine the dependence of the even-$\omega$ and odd-$\omega$ amplitudes on the effective mass of the electrons $m_L$ and chemical potential $\mu_L$ within the nanowire. Throughout this figure, and for much of the results in the remainder of this work, we set the tunneling parameter $t_L=\mu_s/10$, since this leads to a large effect in Fig.~\ref{fig:L+SC_xw}, beyond the weak-tunneling regime.  
From Figs.~\ref{fig:L+SC_x}(a, b), we see that as $m_L$ is increased, the odd-$\omega$ amplitudes are enhanced and the even-$\omega$ amplitudes are slightly suppressed, with the largest effect for both cases appearing around $\omega\approx 0.7\Delta$, as we observed in Fig.~\ref{fig:L+SC_xw}. This enhancement of the odd-$\omega$ amplitude with increasing $m_L$ is consistent with the weak-tunneling results in Eq.~(\ref{eq:FL_odd_approx}), which go as $\sim\sqrt{m_L}$. From Figs.~\ref{fig:L+SC_x}(c, d), we see that as $\mu_L$ is increased, the odd-$\omega$ amplitudes are instead suppressed, while the even-$\omega$ amplitudes are mostly unchanged, though a slight enhancement is just barely noticeable. This suppression of the odd-$\omega$ pairing with large values of $\mu_L$ is also consistent with the weak-tunneling limit in Eq.~(\ref{eq:FL_odd_approx}). Physically, both of these effects can be understood as consequences of the density of states (DOS) in the nanowire, given by $N_L(\omega)=\sqrt{m_L/2(\mu_L+\omega)}/\pi$. Thus when $m_L$ increases, so does the DOS in the nanowire, leading to an enhancement of tunneling processes between the two systems. Likewise, for large values of $\mu_L$ the DOS at low energies decreases, thus suppressing tunneling between the nanowire and the substrate. We therefore see a clear dependence on the induced odd-$\omega$ pair amplitude on the normal state of the nanowire, while the even-$\omega$ pairing is largely unchanged when changing the nanowire as it is dominated by the intrinsic pairing of the superconductor.

To summarize our study of the L+SC system, we note that clearly both even-$\omega$ and odd-$\omega$ pairing are induced in the superconducting substrate. However, only the even-$\omega$ amplitudes exist locally as the odd-$\omega$ amplitude has odd spatial parity. Hence, it is not obvious that these odd-$\omega$ pair amplitudes can have a direct influence on local observables. However, as we will demonstrate in the next section, these odd-$\omega$ pair amplitudes still play a significant role in the local physics when a second nanowire is coupled to the substrate. 

\subsection{Left nanowire + superconductor + right nanowire}
\label{sec:L+SC+R}
The next step in our analysis is to add the right nanowire to the L+SC system studied in detail in the last subsection.
Returning to the Dyson equation for the Green's function of the substrate, Eq.~(\ref{eq:dyson}), and noting that the exact expression for the Green's function of the L+SC system is given by Eq.~(\ref{eq:G_L_explicit}), it is easy to see that the presence of the right nanowire can be accounted for by using:
\begin{equation} 
\hat{\mathcal{G}}_{s}(x_1,x_2)=\hat{\mathcal{G}}^{(L)}_{s}(x_1,x_2)+\hat{\mathcal{G}}^{(L)}_{s}(x_1,\tfrac{d}{2}) \hat{T}_{R}\hat{\mathcal{G}}^{(L)}_{s}(\tfrac{d}{2},x_2).
\label{eq:G_R}
\end{equation}
Here we have defined the $T$-matrix associated with scattering between the L+SC system and the right nanowire as:
\begin{equation}
\hat{T}_{R}=\left[(\hat{\Sigma}^{R}_s)^{-1}- \hat{\mathcal{G}}^{(L)}_{s}(\tfrac{d}{2},\tfrac{d}{2})\right]^{-1},
\label{eq:T_R}
\end{equation}
where $\hat{\mathcal{G}}^{(L)}_{s}(x_1,x_2)$ is given by Eq.~(\ref{eq:G_L_explicit}).
We notice directly that this $T$-matrix for the right nanowire depends only on the local part of the Green's function for the L+SC system, $\hat{\mathcal{G}}^{(L)}_{s}(x,x)$. From Eq.~(\ref{eq:G_L_explicit}) we see that evaluating this Green's function locally simplifies the form somewhat:
\begin{equation}
\begin{aligned}
\hat{\mathcal{G}}^{(L)}_{s}(x,x)=&zg^{(L)}_0(x,x)\hat{\tau}_0+g^{(L)}_3(x,x)\hat{\tau}_3 \\
&+f^{(L)}_1(x,x)\hat{\tau}_1, 
\end{aligned}
\label{eq:localGL}
\end{equation} 
similar to the general form for the bare Green's function of the substrate but now with the coefficients given by Eq.~(\ref{eq:GL0123}). Therefore, using the same reasoning leading to Eq.~(\ref{eq:TL_explicit}), we find that the $T$-matrix capturing the effect of the right nanowire is given by: 
\begin{equation}
\hat{T}_{R}=zT^{(R)}_0\hat{\tau}_0+T^{(R)}_1\hat{\tau}_1+T^{(R)}_3\hat{\tau}_3
\label{eq:T_R_explicit}
\end{equation}
where the coefficients $T^{(R)}_0$, $T^{(R)}_1$, and $T^{(R)}_3$ are given in Eq.~(\ref{eq:TR_123}), and possess the same symmetries as the coefficients in Eq.~(\ref{eq:TL_explicit}) for the L+SC system. 

Since Eq. (\ref{eq:T_R_explicit}) has the same form and symmetries as Eq.~(\ref{eq:TL_explicit}) for the L+SC system, we find that additional pair symmetry conversion must occur in the presence of the right nanowire. Still, since the nanowires are both trivial in the spin index, the induced odd-$\omega$ pairing is always odd in parity and, thus, inherently non-local. However, in contrast to the single nanowire results in the L+SC system in Sec.~\ref{sec:L+SC}, the additional right nanowire allows the non-local correlations of the L+SC system, and in particular the odd-$\omega$ components, to directly influence the local correlations of the total L+SC+R system. Thus there is still a local physical effect of odd-$\omega$ pairing in the presence of the two nanowires.

To make this statement about the influence of the odd-$\omega$ pairing more concrete we explicitly calculate the local Green's function of the L+SC+R system $\hat{\mathcal{G}}_{s}(x,x)$ in terms of the Green's function of the L+SC system $\hat{\mathcal{G}}^{(L)}_{s}(x_1,x_2)$, using Eq.~(\ref{eq:G_R}). We find that $\hat{\mathcal{G}}_{s}(x,x)$ is given by:
\begin{equation}
\begin{aligned}
\hat{\mathcal{G}}_{s}(x,x)=&zg^{(R)}_0(x)\hat{\tau}_0+g^{(R)}_3(x)\hat{\tau}_3+f^{(R)}_1(x)\hat{\tau}_1, 
\end{aligned}
\label{eq:localGR}
\end{equation} 
where the coefficients $g^{(R)}_0$, $g^{(R)}_3$, and $f^{(R)}_1$ are given in Eq.~(\ref{eq:GR0123}) in the Appendix. Examining these expressions, we notice that all three components depend directly on the non-local odd-$\omega$ pair amplitude of the L+SC system, $f^{(L)}_2(x,\tfrac{d}{2})$. Thus, clearly, the odd-$\omega$ pairing influences local properties in the double nanowire system, even though it has an odd spatial parity. 

We now focus on the influence of the odd-$\omega$ terms in the L+SC system on the local pair amplitudes of the full double nanowire system L+SC+R.  For this we decompose the local pair amplitude, $F_s(x)=f^{(R)}_1(x)$, in the following way:
\begin{equation}
F_s(x)= F_{e}(x) + F_{o}(x),
\label{eq:F_local}
\end{equation}
where we define
\begin{widetext}
\begin{equation}
\begin{aligned}
F_{e}(x)=& f^{(L)}_1(x,x)+T^{(R)}_1\left[z^2g^{(L)}_0(x,\tfrac{d}{2})^2-g^{(L)}_3(x,\tfrac{d}{2})^2+f^{(L)}_1(x,\tfrac{d}{2})^2\right] + 2T^{(R)}_3g^{(L)}_3(x,\tfrac{d}{2})f^{(L)}_1(x,\tfrac{d}{2}) \\
&+  2z^2T^{(R)}_0g^{(L)}_0(x,\tfrac{d}{2})f^{(L)}_1(x,\tfrac{d}{2}), \\
F_{o}(x)=&-z^2f^{(L)}_2(x,\tfrac{d}{2})\left[2T_0^{(R)}g^{(L)}_3(x,\tfrac{d}{2})+T_1^{(R)}f^{(L)}_2(x,\tfrac{d}{2}) +2T_3^{(R)}g^{(L)}_0(x,\tfrac{d}{2})\right]. \\
\end{aligned}
\label{eq:feo}
\end{equation}
\end{widetext}
Here, $F_e$ depends only on the even-$\omega$ pair amplitudes of the L+SC system, $f_1^{(L)}$, or its normal state, while $F_o$ gathers all terms that depends directly on the the odd-$\omega$ pair amplitudes of the L+SC system, $f_2^{(L)}$. We emphasize that the local pair amplitude in Eq.~(\ref{eq:F_local}), $F_s(x)$, is purely even-$\omega$, and yet, through higher-order scattering processes it has acquired a dependence on the non-local odd-$\omega$ pair amplitudes of the L+SC system, which we write as $F_{o}(x)$ in Eq. (\ref{eq:feo}).
 
\begin{figure}[htb]
 \begin{center}
  \centering
        \includegraphics[width=0.5\textwidth]{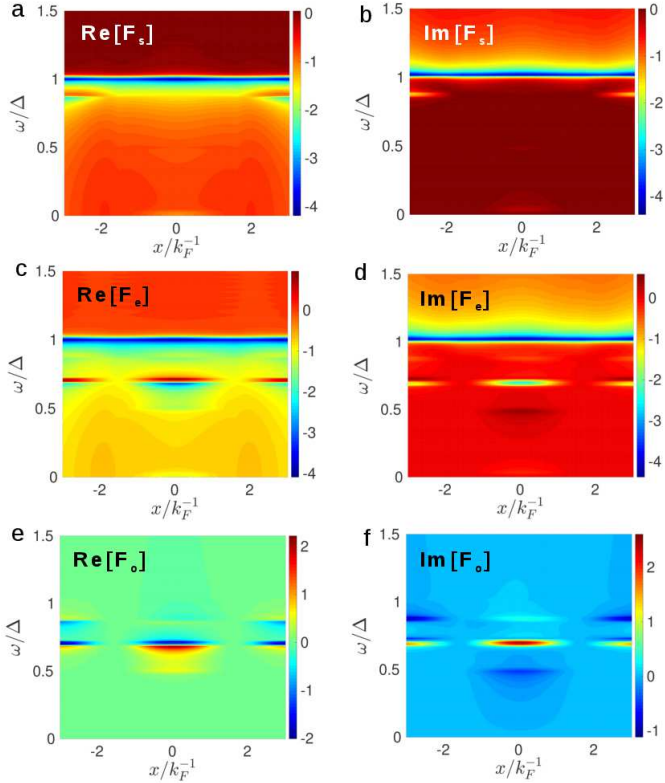}
  \caption{Density plots in the $x,\omega$-plane of: real (a) and imaginary (b) parts of the retarded local pair amplitude of the superconducting substrate for the L+SC+R system, $F_s(x,z=\omega+i\eta)$, given by Eq.~(\ref{eq:F_local}); real (c) and imaginary (d) parts of the even-$\omega$ contributions, $F_{e}(x,z=\omega+i\eta)$, given by Eq.~(\ref{eq:feo}); and real (e) and imaginary (f) parts of the odd-$\omega$ contributions, $F_{o}(x)$, given by Eq.~(\ref{eq:feo}). We use the same units and parameters as in Fig.~\ref{fig:L+SC_xw}(e,f): $\mu_L=\mu_R=\Delta/2$, $m_L=m_R=m_s$, and $d=\pi k_F^{-1}$, where $k_F=\sqrt{2m_s\mu_s}$, $\Delta=\mu_s/100$, and $t_L=t_R=\mu_s/10$.}
  \label{fig:L+SC+R_xw}
 \end{center}
\end{figure}

To gain insight into the behavior of the even- and odd-$\omega$ origins of $F_s$, we plot the real and imaginary parts of the retarded versions of $F_s(x,z=\omega+i\eta)$, $F_{e}(x,z=\omega+i\eta)$, and $F_{o}(x,z=\omega+i\eta)$ in-between the two wires in Fig.~\ref{fig:L+SC+R_xw}. We choose parameters such that there is appreciable odd-$\omega$ pairing in the L+SC system according to our results in the previous subsection. 
In Figs.~\ref{fig:L+SC+R_xw} (a, b), we notice that the total local pair amplitude, given by Eq.~(\ref{eq:F_local}), possesses most of its weight around the gap, i.e.~$\omega\approx\Delta$, as expected. It is also not rapidly varying with respect to $x$ in this range. Comparing this result to contributions coming from the even-$\omega$ amplitude of the L+SC system, Figs.~\ref{fig:L+SC+R_xw} (c, d), we notice one major difference: $F_e$ possesses strong features around $\omega\approx 0.7\Delta$, the frequency associated with the peaks in the odd-$\omega$ pair amplitude shown in Figs.~\ref{fig:L+SC_xw} (e,f). Turning our attention to the contributions coming from the odd-$\omega$ amplitudes of the L+SC system, Figs.~\ref{fig:L+SC+R_xw} (e, f), we see that these notable features in $F_e$ are smoothed out by features with the opposite signs coming from the odd-$\omega$ amplitude $F_o$. This demonstrates that the non-local odd-$\omega$ pair amplitude of the L+SC system has an appreciable effect on the local pair amplitude of the L+SC+R system. Furthermore, we can understand the $x$-dependence of these features by noting that $F_o$ is proportional to $f_2^{(L)}(x,\tfrac{d}{2})$ in Eq. (\ref{eq:feo}). From the weak-tunneling expression in Eq. (\ref{eq:FL_odd_approx}) we can see that $f_2^{(L)}(x,\tfrac{d}{2})$ is maximized when $2k_F(|x+d/2|-|d|)=(2n+1)\pi$ which is exactly what we observe in Fig.~\ref{fig:L+SC+R_xw}.

\begin{figure}[htb]
 \begin{center}
  \centering
        \includegraphics[width=0.45\textwidth]{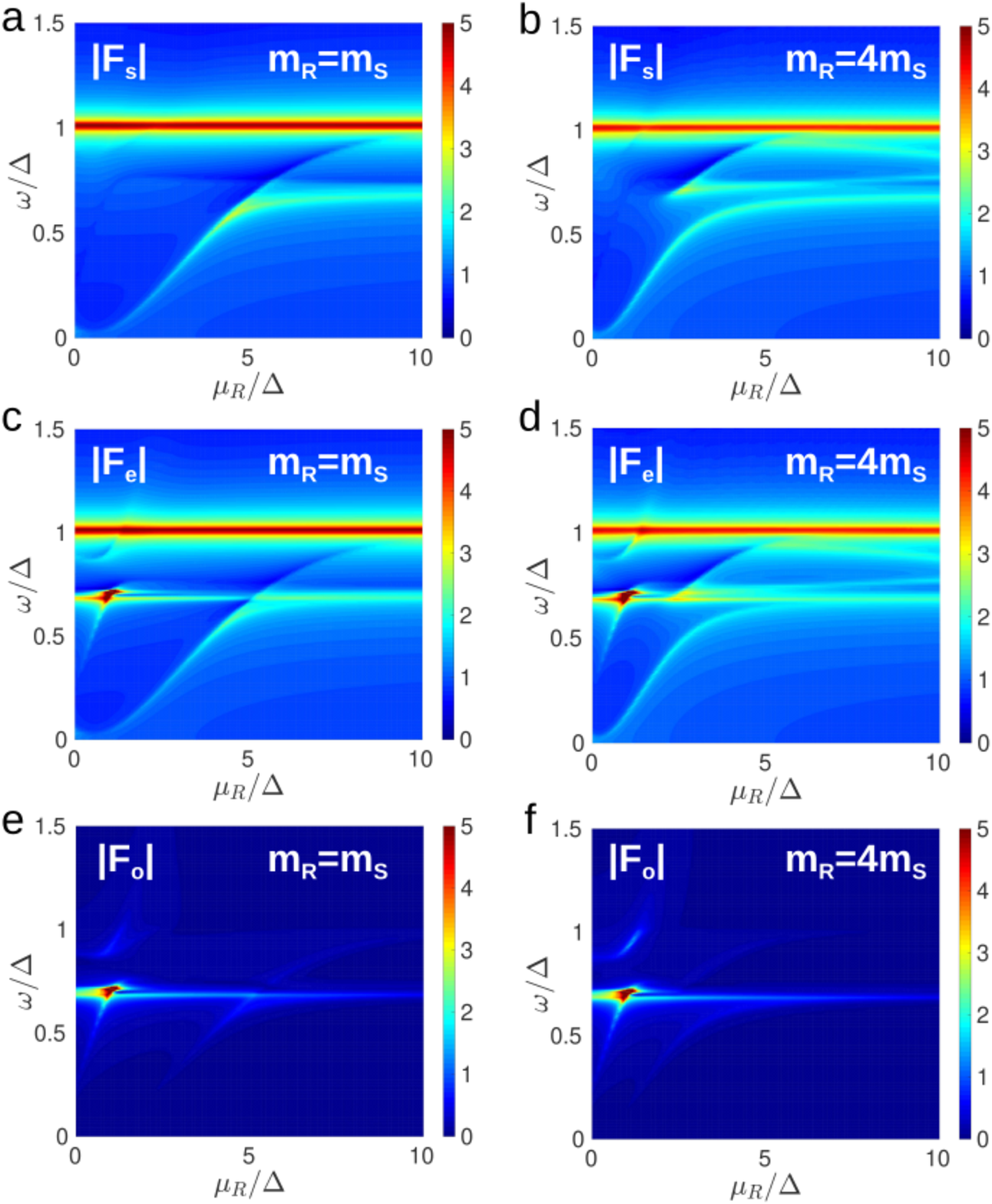}
  \caption{Density plots in the $\mu_R,\omega$-plane of the absolute magnitude of: $|F_s(x=0,\omega+i\eta)|$, given by Eq.~(\ref{eq:F_local}), (a) and (b); $|F_{e}(x=0,\omega+i\eta)|$, given by Eq.~(\ref{eq:feo}), (c) and (d); and $|F_{o}(x=0,\omega+i\eta)|$, given by Eq.~(\ref{eq:feo}), (e) and (f). In each case, we have fixed the model parameters such that $m_L=m_s$, $\mu_L=\Delta/2$, $\Delta=\mu_s/100$, $d=\pi k_F^{-1}$, where $k_F=\sqrt{2m_s\mu_s}$, and plotted each function for two different values of $m_R$, as indicated in each panel.}
  \label{fig:L+SC+R_wmur}
 \end{center}
\end{figure}

\begin{figure}[htb]
 \begin{center}
  \centering
        \includegraphics[width=0.45\textwidth]{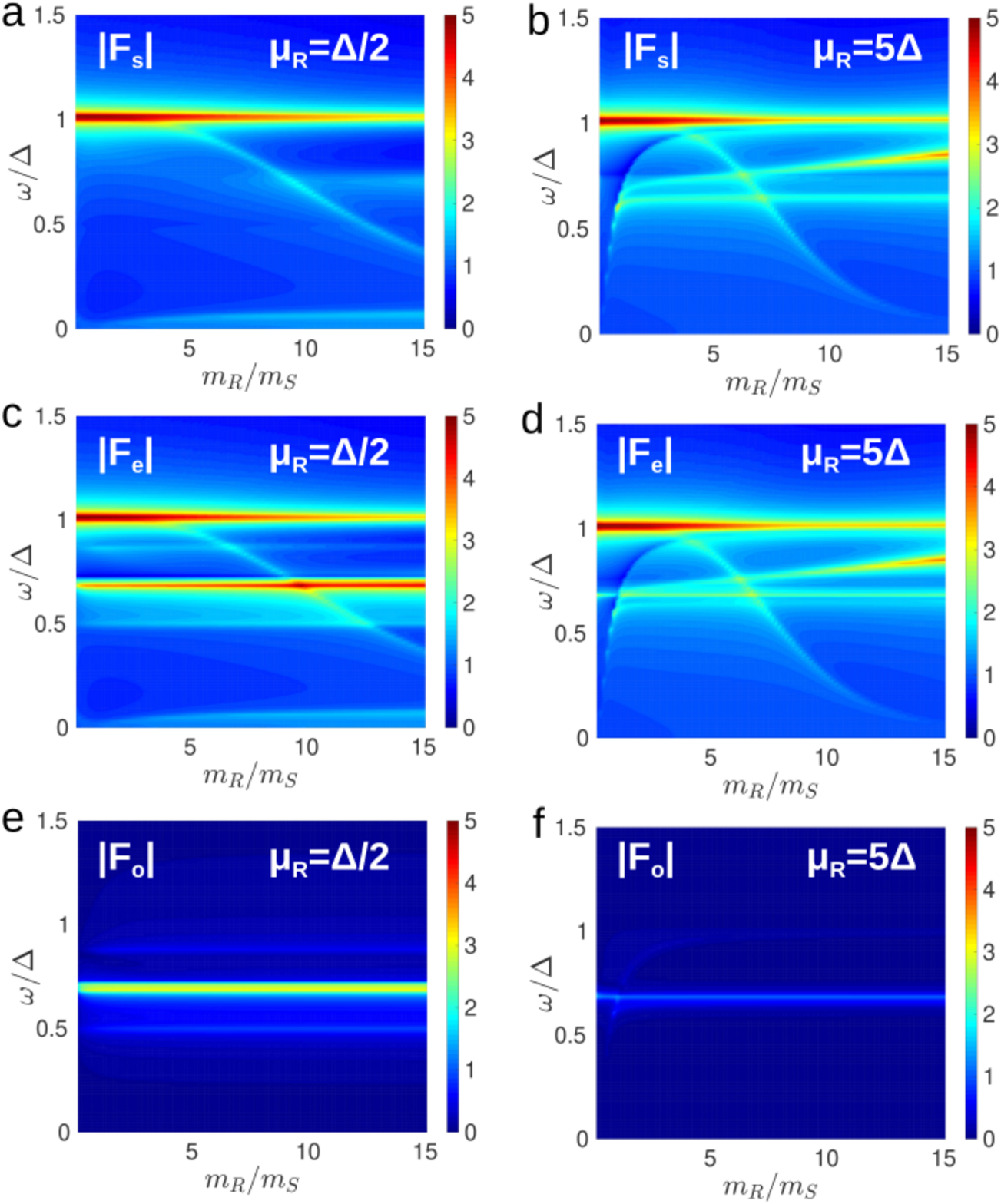}
  \caption{Density plots in the $m_R,\omega$-plane of the absolute magnitude of: $|F_s(x=0,\omega+i\eta)|$, given by Eq.~(\ref{eq:F_local}), (a) and (b); $|F_{e}(x=0,\omega+i\eta)|$, given by Eq.~(\ref{eq:feo}), (c) and (d); and $|F_{o}(x=0,\omega+i\eta)|$, given by Eq.~(\ref{eq:feo}), (e) and (f). In each case, we have fixed the model parameters such that $m_L=m_s$, $\mu_L=\Delta/2$, $\Delta=\mu_s/100$, $d=\pi k_F^{-1}$, where $k_F=\sqrt{2m_s\mu_s}$, and plotted each function for two different values of $\mu_R$, as indicated in each panel.}
  \label{fig:L+SC+R_wmr}
 \end{center}
\end{figure}

\begin{figure}[htb]
 \begin{center}
  \centering
        \includegraphics[width=0.45\textwidth]{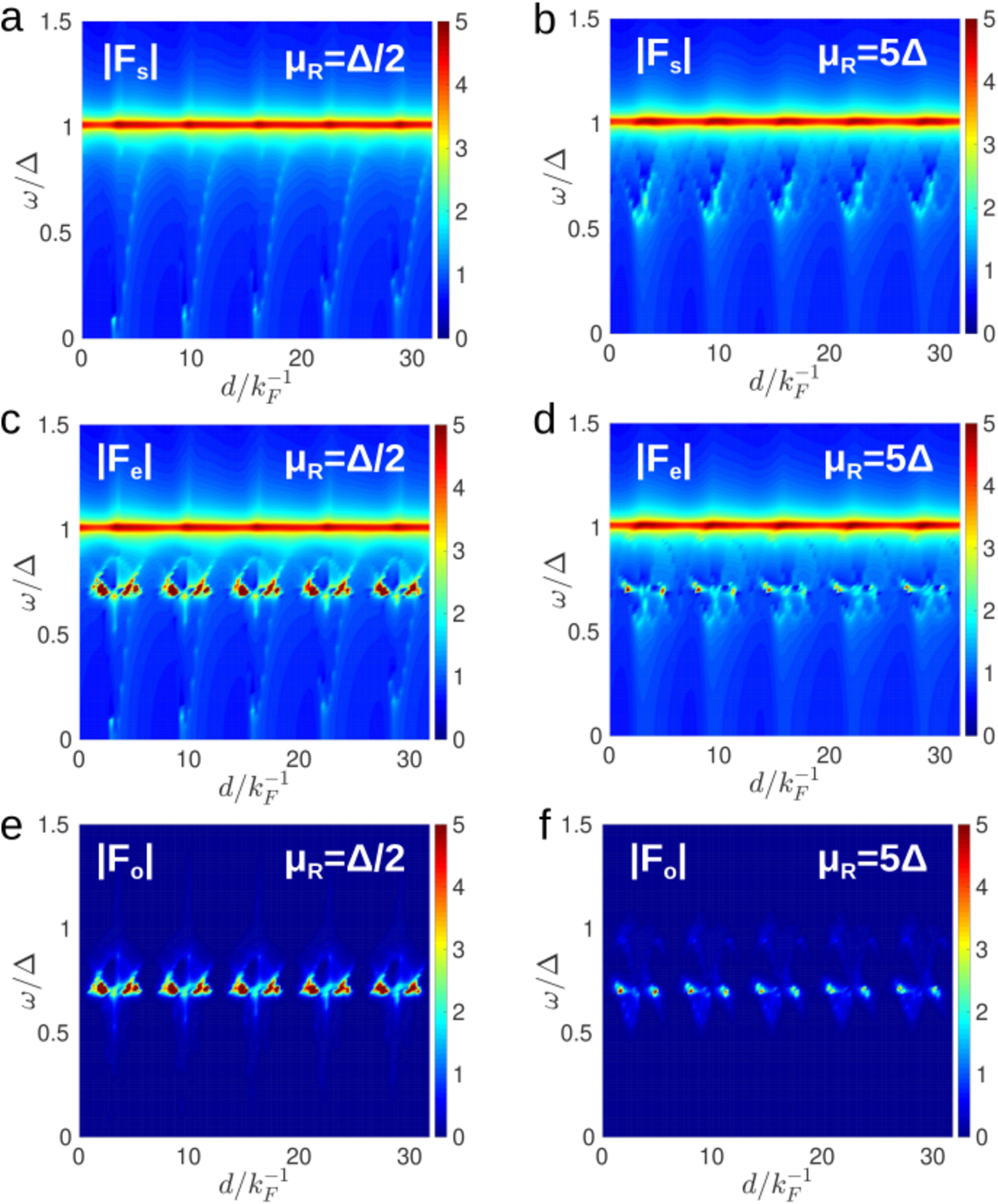}
  \caption{Density plots in the $d,\omega$-plane of the absolute magnitude of: $|F_s(x=0,\omega+i\eta)|$, given by Eq.~(\ref{eq:F_local}), (a) and (b); $|F_{e}(x=0,\omega+i\eta)|$, given by Eq.~(\ref{eq:feo}), (c) and (d); and $|F_{o}(x=0,\omega+i\eta)|$, given by Eq.~(\ref{eq:feo}), (e) and (f). In each case, we have fixed the model parameters such that $m_R=m_L=m_s$, $\mu_L=\Delta/2$, $\Delta=\mu_s/100$, distances are given in units of $k_F^{-1}=1/\sqrt{2m_s\mu_s}$, and plotted each function for two different values of $\mu_R$, as indicated in each panel.}
  \label{fig:L+SC+R_wd}
 \end{center}
\end{figure}

In Fig.~\ref{fig:L+SC+R_wmur} we present density plots of the absolute magnitudes of $F_s$, $F_e$, and $F_o$ as a function of $\omega$ and $\mu_R$, at the position $x=0$, the position at which the contribution from $F_o$ is most pronounced according to Fig.~\ref{fig:L+SC+R_xw}. In these plots we have fixed the parameters of the left nanowire to the same as Fig.~\ref{fig:L+SC_xw}(e,f) and Fig.~\ref{fig:L+SC+R_xw}: $m_L=m_s$, $\mu_L=\Delta/2$; and we have chosen a value for the interwire separation associated with a sizable contribution from $f_2^{(L)}$, $d=\pi k_F^{-1}$. In the left column, Fig.~\ref{fig:L+SC+R_wmur} (a,c,e), we show the results for an effective mass of $m_R=m_L=m_s$, while in the right column, Fig.~\ref{fig:L+SC+R_wmur} (b,d,f), we show the results for an effective mass of $m_R=4m_L=4m_s$. Notice that both values of $m_R$ behave similarly and that there is a sizable contribution from the odd-$\omega$ pairing in the region where $\omega\approx 0.7\Delta$ and $\mu_R\leq\Delta$. This contribution appears to decay rapidly for $\mu_R>\Delta$, and is cancelled, for the most part, by a contribution from the even-$\omega$ amplitude. However, we note that in both cases, the contribution from the even-$\omega$ pairing around $\omega\approx 0.7\Delta$ does not fully decay for large $\mu_R$. Instead, this contribution leads to a feature in the total pair amplitude which remains pinned to $\omega\approx 0.7\Delta$. The precise frequency at which we find this feature depends on the parameters of the substrate; however, we note that it occurs at the same frequency as the peak structures observed in $f_2^{(L)}$ which were discussed the previous subsection. In the case of $F_o$, this relationship is direct, since $F_o$ is proportional to $f_2^{(L)}$ and therefore shares much of its peak structure. From Eq. (\ref{eq:FL_odd_approx}), we see that in the weak-tunneling limit, these features of $f_2^{(L)}$ will occur at $\omega=\pm\mu_L,\pm\Delta$, while for the stronger-tunneling value considered here, $t_L=\mu_s/10$, we find these frequencies pinned just below the gap, around $\omega\approx 0.7\Delta$.

To better understand how the features we observed in Fig.~\ref{fig:L+SC+R_wmur} depend on the other model parameters, specifically the effective mass in the right nanowire, $m_R$, and the interwire separation, $d$, we present in Figs.~\ref{fig:L+SC+R_wmr} and \ref{fig:L+SC+R_wd} similar density plots to the ones shown in Fig.~\ref{fig:L+SC+R_wmur}, but in the $m_R,\omega$- and $d,\omega$- planes, respectively. In both Figs.~\ref{fig:L+SC+R_wmr} and \ref{fig:L+SC+R_wd}, we can clearly see that the main contribution from the odd-$\omega$ pair amplitude keeps occuring at $\omega\approx 0.7\Delta$, and that it is maximized when $\mu_R\leq\Delta$, consistent with the behavior observed in Fig.~\ref{fig:L+SC+R_wmur}.

Focusing on Fig.~\ref{fig:L+SC+R_wmr}, we see that, for $\mu_R=\Delta/2$ the contribution from the odd-$\omega$ pair amplitude, $F_o$, is comparable in magnitude to the largest contribution from the even-$\omega$ pair amplitude, $F_e$, but the two contributions seem to cancel over the entire range of $m_R$ considered. Furthermore, we note that, as $m_R$ is increased these contributions quickly reach a plateau and appear to remain more or less constant for larger values of $m_R$. However, for $\mu_R=5\Delta$ we find that $F_o$ is significantly smaller than $F_e$. This decrease in the magnitude of $F_o$ is accompanied by the emergence of a feature in the total local pair amplitude for much of the range considered, as we also found in Fig.~\ref{fig:L+SC+R_wmur}. 

Finally, turning our attention to Fig.~\ref{fig:L+SC+R_wd}, we confirm that $F_o$ possesses a strong peak around $\omega\approx 0.7\Delta$, as we observed in Figs.~\ref{fig:L+SC+R_wmur} and \ref{fig:L+SC+R_wmr}, and that this contribution and $F_e$ exactly cancel for $\mu_R=\Delta/2$, but that the cancellation does not occur for $\mu_R=5\Delta$, leading to the appearance of features in the total pair amplitude at $\omega\approx 0.7\Delta$. Additionally, in Fig.~\ref{fig:L+SC+R_wd} we see that all three amplitudes, $F_s$, $F_e$, and $F_o$, have an approximately periodic dependence on the interwire separation, $d$, with period roughly $\sim 2\pi k_F^{-1}$. In particular, we see that $F_o$ acquires its largest magnitude at values of the interwire separation associated with odd integer multiples of $\pi/2$, $d\approx (2n+1)\pi/2 k_F^{-1}$, as also seen in the sinusoidal behavior of $f_2^{(L)}$ in Eq. (\ref{eq:FL_odd_approx}).

\section{Local Experimental Signatures of Non-Local Odd-frequency Amplitudes}
\label{sec:obs}
In the previous section we demonstrated that the non-local odd-$\omega$ pair amplitudes have a direct effect on the local even-$\omega$ pair amplitudes. As such, there should be signatures of odd-$\omega$ pairing in local, experimentally-observable, quantities. In this section we show that this is indeed the case for two easily measurable observables, the LDOS and local Josephson current.

\subsection{Local density of states}
We compute the LDOS of quasiparticles in the usual way, using the retarded Green's function:
\begin{equation}
\mathcal{N}(\textbf{R};\omega)=-\frac{1}{\pi}\text{Im}\text{Tr}\hat{\mathcal{G}}(\textbf{R},\textbf{R};\omega).
\label{eq:ldos}
\end{equation} 
This quantity can be measured experimentally through STM using a well-characterized normal metal tip. 
From Eqs.~(\ref{eq:localGR}) and (\ref{eq:ldos}) we find that the LDOS for the superconducting substrate coupled to both nanowires takes the form:
\begin{equation}
\mathcal{N}(x;\omega)=-\frac{2}{\pi}\text{Im}\left[(\omega+i\eta) \int_{-\infty}^{\infty}\frac{dk_y}{2\pi} g^{(R)}_0(x,k_y,\omega) \right],
\label{eq:ldos_total}
\end{equation}
where $g^{(R)}_0$ is given by Eq.~(\ref{eq:GR0123}) and we have set $y=0$ for simplicity. By inspecting Eq.~(\ref{eq:GR0123}), we can also isolate the contribution to the LDOS arising from the odd-$\omega$ pair amplitudes:
\begin{widetext}
\begin{equation}
\begin{aligned}
\mathcal{N}_{o}(x;\omega)=&-\frac{2}{\pi}\text{Im}\left\{(\omega+i\eta) \int_{-\infty}^{\infty}\frac{dk_y}{2\pi}f^{(L)}_2(x,\tfrac{d}{2}) \left[ T^{(R)}_0(\omega+i\eta)^2f^{(L)}_2(x,\tfrac{d}{2})+2T^{(R)}_1g^{(L)}_3(x,\tfrac{d}{2}) -2T^{(R)}_3f^{(L)}_1(x,\tfrac{d}{2}) \right] \right\},
\end{aligned}
\label{eq:ldos_odd}
\end{equation}
which is clearly proportional to the non-local odd-$\omega$ pair amplitude generated by the left nanowire, $f^{(L)}_2$. For completeness we also write out the whole contribution to the LDOS which does not depend on the odd-$\omega$ pair amplitudes, $\mathcal{N}_e\equiv \mathcal{N}- \mathcal{N}_o$, given by:
\begin{equation}
\begin{aligned}
\mathcal{N}_{e}(x;\omega)=&-\frac{2}{\pi}\text{Im}\left\{(\omega+i\eta) \int_{-\infty}^{\infty}\frac{dk_y}{2\pi}\left[ g^{(L)}_0(x,x)+T^{(R)}_0\left((\omega+i\eta)^2g^{(L)}_0(x,\tfrac{d}{2})^2+g^{(L)}_3(x,\tfrac{d}{2})^2+f^{(L)}_1(x,\tfrac{d}{2})^2\right) \right.\right. \\
&+\left. \left. 2T^{(R)}_1 g^{(L)}_0(x,\tfrac{d}{2})f^{(L)}_1(x,\tfrac{d}{2}) +2T^{(R)}_3g^{(L)}_0(x,\tfrac{d}{2})g^{(L)}_3(x,\tfrac{d}{2}) \right] \right\},
\end{aligned}
\label{eq:ldos_notodd}
\end{equation}
\end{widetext}  
which instead depends only on the even-$\omega$ amplitude $f^{(L)}_1$ and the normal quasiparticle terms $g^{(L)}_0$ and $g^{(L)}_3$. We note that, in real experiments it is the total LDOS, Eq. (\ref{eq:ldos_total}), that is measured, not the separate contributions given by Eqs. (\ref{eq:ldos_odd}) and (\ref{eq:ldos_notodd}). However, it is instructive to examine these separate terms to understand which features in the total LDOS are directly influenced by the odd-$\omega$ pairing and which are not.

\begin{figure}[htb]
 \begin{center}
  \centering
        \includegraphics[width=0.45\textwidth]{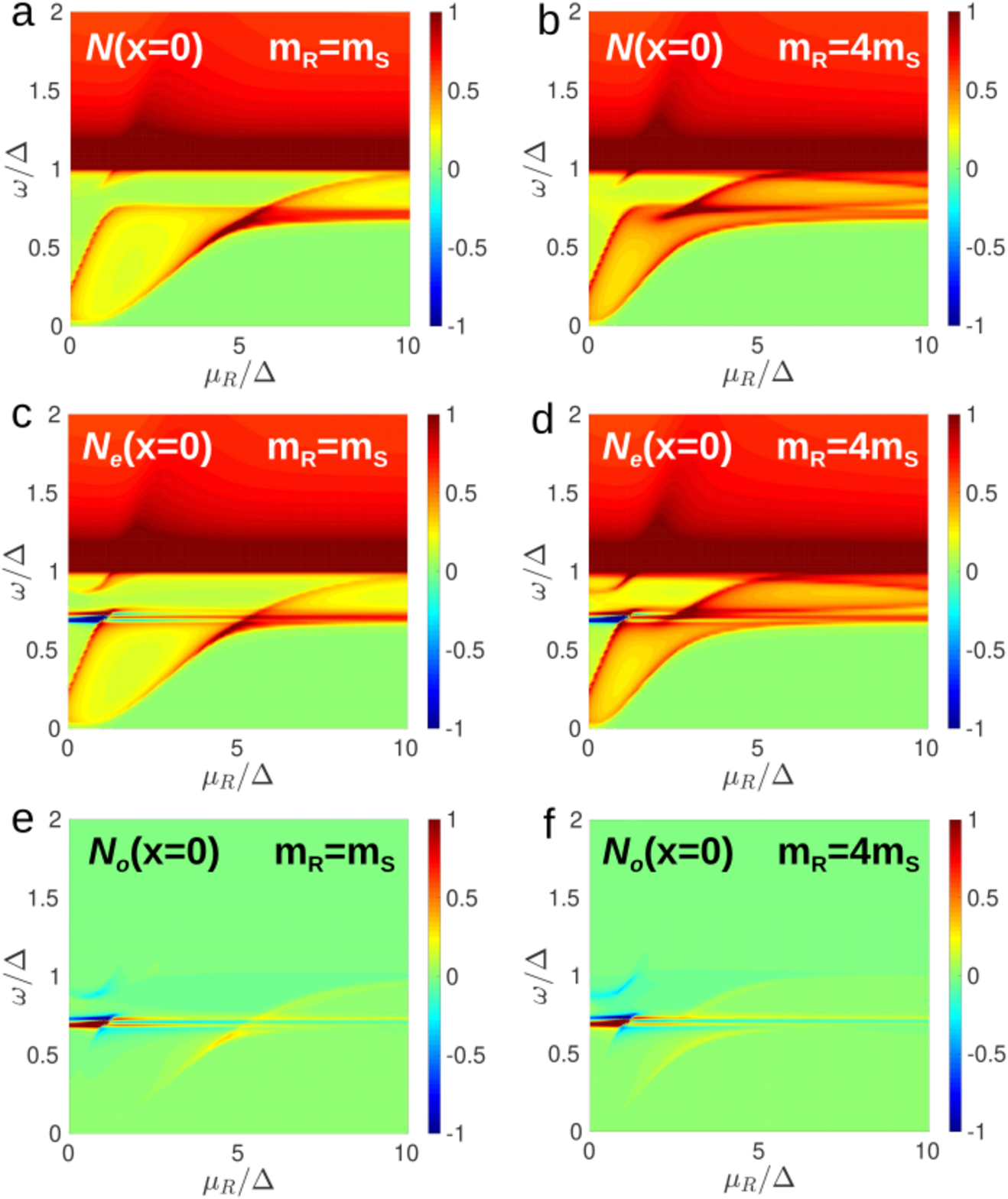}
  \caption{Density plots of the contributions to the LDOS in the $\mu_R,\omega$-plane. (a) and (b) the total LDOS, $\mathcal{N}(x=0;\omega)$, given by Eq.~(\ref{eq:ldos_total}); (c) and (d) the even-$\omega$ contribution, $\mathcal{N}_{e}(x=0;\omega)$, given by Eq.~(\ref{eq:ldos_notodd}); and (e) and (f) the odd-$\omega$ contribution, $\mathcal{N}_{o}(x=0;\omega)$, given by Eq.~(\ref{eq:ldos_odd}). In each case, we have fixed the model parameters to the same values as Fig.~\ref{fig:L+SC+R_wmur}.}
  \label{fig:ldos_L+SC+R_wmur}
 \end{center}
\end{figure}

\begin{figure}
 \begin{center}
  \centering
        \includegraphics[width=0.45\textwidth]{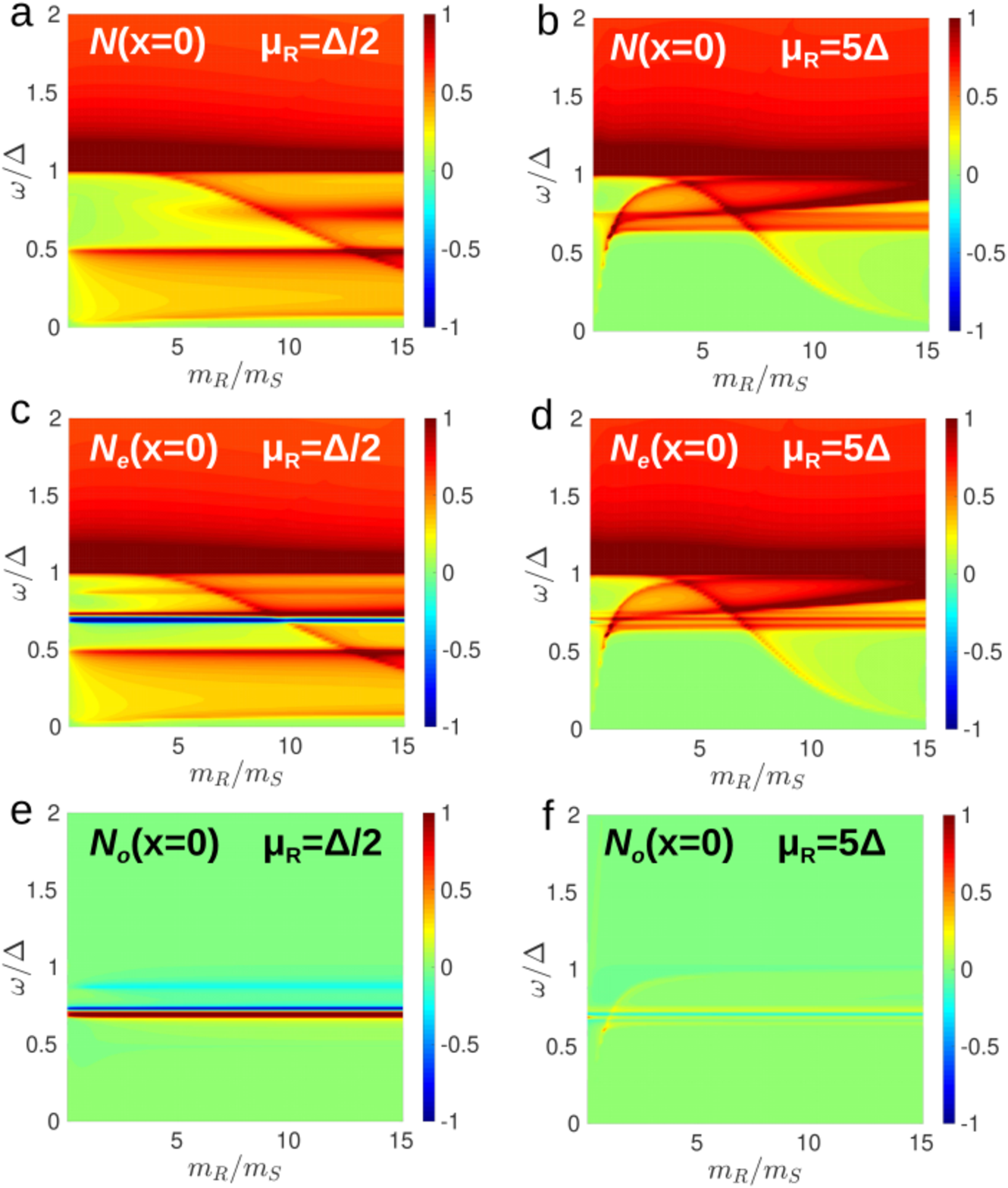}
  \caption{Density plots of the contributions to the LDOS in the $m_R,\omega$-plane. (a) and (b) the total LDOS, $\mathcal{N}(x=0;\omega)$, given by Eq.~(\ref{eq:ldos_total}); (c) and (d) the even-$\omega$ contribution, $\mathcal{N}_{e}(x=0;\omega)$, given by Eq.~(\ref{eq:ldos_notodd}); and (e) and (f) the odd-$\omega$ contribution, $\mathcal{N}_{o}(x=0;\omega)$, given by Eq.~(\ref{eq:ldos_odd}). In each case, we have fixed the model parameters to the same values as Fig.~\ref{fig:L+SC+R_wmr}.}
  \label{fig:ldos_L+SC+R_wmr}
 \end{center}
\end{figure}

\begin{figure}
 \begin{center}
  \centering
        \includegraphics[width=0.45\textwidth]{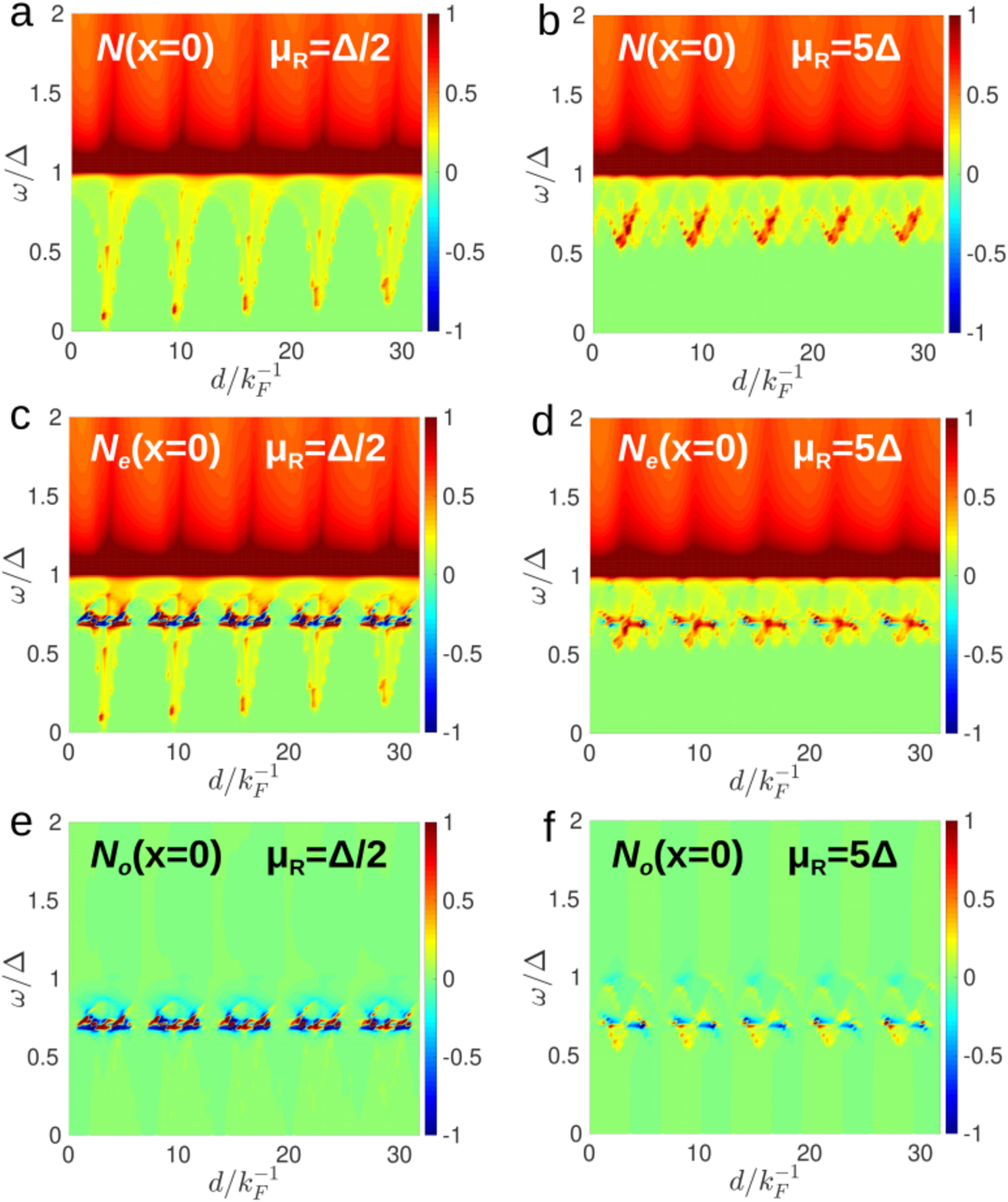}
  \caption{Density plots of the contributions to the LDOS in the $d,\omega$-plane. (a) and (b) the total LDOS, $\mathcal{N}(x=0;\omega)$, given by Eq.~(\ref{eq:ldos_total}); (c) and (d) the even-$\omega$ contribution, $\mathcal{N}_{e}(x=0;\omega)$, given by Eq.~(\ref{eq:ldos_notodd}); and (e) and (f) the odd-$\omega$ contribution, $\mathcal{N}_{o}(x=0;\omega)$, given by Eq.~(\ref{eq:ldos_odd}). In each case, we have fixed the model parameters to the same values as Fig.~\ref{fig:L+SC+R_wd}.}
  \label{fig:ldos_L+SC+R_wd}
 \end{center}
\end{figure}

To proceed we have numerically evaluated all three contributions: the total LDOS, given by Eq. (\ref{eq:ldos_total}); the contribution coming from the odd-$\omega$ pairing, Eq. (\ref{eq:ldos_odd}); and the remaining contribution, $\mathcal{N}_e$, given by Eq. (\ref{eq:ldos_notodd}). To understand how these contributions depend on the different model parameters we present density plots of these expressions in Figs. \ref{fig:ldos_L+SC+R_wmur}-\ref{fig:ldos_L+SC+R_wd} for the same parameters we used to study the pair amplitudes in Figs. \ref{fig:L+SC+R_wmur}-\ref{fig:L+SC+R_wd}, focusing in each case on the behavior at $x=0$, i.e. the midpoint between the two wires, where the odd-$\omega$ pairing have the largest impact on the local pair amplitude according to the previous section.

In Fig.~\ref{fig:ldos_L+SC+R_wmur} we show the results for $\mathcal{N}$, $\mathcal{N}_e$, and $\mathcal{N}_o$, respectively, in the $\mu_R,\omega$-plane, for the same parameters as the corresponding plots of the pair amplitudes in Fig.~\ref{fig:L+SC+R_wmur}. As with the pair amplitudes, we notice that the largest contribution from the odd-$\omega$ pairing appears for small $\mu_R$ relative to $\Delta$ and is pinned to the frequency $\omega \approx 0.7\Delta$. Also, similar to the pair amplitudes in Fig.~\ref{fig:L+SC+R_wmur}, as $\mu_R$ increases we see that the odd-$\omega$ contribution decreases while the $\mathcal{N}_e$ contribution retains its spectral weight near $\omega \approx 0.7\Delta$ so that for large $\mu_R$ this feature is observable in the total LDOS, $\mathcal{N}$. In contrast to the pair amplitudes, the $\mathcal{N}$ and $\mathcal{N}_e$ contributions both possess noticeably more spectral weight below the gap, especially at smaller values of $\mu_R$.         

In Fig. \ref{fig:ldos_L+SC+R_wmr} we show similar LDOS results to those appearing in Fig. \ref{fig:ldos_L+SC+R_wmur} but in the $m_R,\omega$-plane, and using the same parameters as the corresponding plots of the pair amplitudes in Fig.~\ref{fig:L+SC+R_wmr}. In Figs. \ref{fig:ldos_L+SC+R_wmr} (a,c,e) we set $\mu_R=\Delta/2$, while for Figs. \ref{fig:ldos_L+SC+R_wmr} (b,d,f) we set $\mu_R=5\Delta$. Once again we find that the largest contribution from the odd-$\omega$ pairing appears for the smaller value of $\mu_R$ and with essentially all of its spectral weight near $\omega \approx 0.7\Delta$, possessing parameter dependences similar to the pair amplitudes in Fig.~\ref{fig:L+SC+R_wmr}. However, in addition to the feature pinned to $\omega\approx 0.7\Delta$, the $\mathcal{N}_e$ and $\mathcal{N}$ contributions possess noticeably larger subgap features when $\mu_R=\Delta/2$ in contrast to the results for the pair amplitudes, but consistent with the behavior of the LDOS in Fig.~\ref{fig:ldos_L+SC+R_wmur}.   

Finally, in Fig. \ref{fig:ldos_L+SC+R_wd} we show similar LDOS results to those appearing in Figs. \ref{fig:ldos_L+SC+R_wmur} and \ref{fig:ldos_L+SC+R_wmr} but in the $d,\omega$-plane, plotted for the same parameters as the pair amplitudes appearing in Fig.~\ref{fig:L+SC+R_wd}. In Figs. \ref{fig:ldos_L+SC+R_wd} (a,c,e) we set $\mu_R=\Delta/2$, while for Figs. \ref{fig:ldos_L+SC+R_wd} (b,d,f) we set $\mu_R=5\Delta$. Similar to the results for the pair amplitudes in Fig. \ref{fig:L+SC+R_wd}, we find that all subgap contributions possess a $d$-dependence, with the most significant subgap features appearing at $d=(2n+1)\pi k_F^{-1}$, the same values at which the odd-$\omega$ amplitude, $f_2^{(L)}$, reaches its largest magnitudes. 

Comparing Figs.~\ref{fig:ldos_L+SC+R_wmur}-\ref{fig:ldos_L+SC+R_wd}, we see that, overall, they possess similar parameter dependences to the corresponding plots of the pair amplitudes in Figs.~\ref{fig:L+SC+R_wmur}-\ref{fig:L+SC+R_wd}, but with some notable differences. Similar to the local pair amplitudes, the LDOS has contributions coming directly from the odd-$\omega$ pair amplitudes, and these features are primarily pinned to the same value of $\omega$ as the contributions to the local pair amplitudes. Also, similar to the pair amplitudes, the effect of the odd-$\omega$ pairing on the LDOS is such that it often cancels a corresponding spectral weight from $\mathcal{N}_e$, so that the total LDOS $\mathcal{N}$ remains featureless. However, since $\mathcal{N}_o$ and $\mathcal{N}_e$ have different parameter dependence, when the contribution from $\mathcal{N}_o$ is suppressed these features show up in the total LDOS. On the other hand, one important difference between the pair amplitudes and the LDOS is that $\mathcal{N}$ and $\mathcal{N}_e$ both possess strong subgap peaks at low values of $\mu_R$, which do not appear to be related to the odd-$\omega$ pair amplitudes. Since these features only appear in $\mathcal{N}$ and $\mathcal{N}_e$, they are likely due to the normal quasiparticles in the system. Therefore, while the LDOS does exhibit features attributable to the odd-$\omega$ pairing, a more direct probe of the local Cooper pairs could display more clear signatures of these odd-$\omega$ pair amplitudes.

\subsection{Josephson tunneling spectroscopy}

While the LDOS can be accessed by STM using a well-characterized normal metal tip, by using a superconducting tip, similar measurements can probe the local Cooper pair superfluid density, directly.\cite{naaman2001fluctuation,vsmakov2001josephson,rodrigo2004use,yokoyama2008theory,hamidian2016detection,randeria2016scanning} Such experiments measure the local Josephson current (LJC) between a superconducting tip and the superconducting substrate. In the limit of weak tunneling between the tip and the substrate, and at zero bias, the LJC at position $\textbf{r}_0$ is given by\cite{mahan2013many}:
\begin{equation}
\begin{aligned}
I_{J}(\textbf{r}_0)=4et_0^2 T \sum_n&\text{Im}\left\{e^{i\Phi} \bar{F}_{tip}(i\omega_n) F_{s}(\textbf{r}_0,\textbf{r}_0;i\omega_n)  \right\}, \\
\end{aligned}
\label{eq:ljc}
\end{equation}
where $F_{tip}(i\omega_n)$ and $F_{s}(\textbf{r}_0,\textbf{r}_0;i\omega_n)$ are the anomalous Matsubara Green's functions of the superconducting tip and superconducting substrate, respectively. Moreover, $\Phi$ is the difference between the phase of the order parameters in the tip and the substrate, $t_0$ is the hopping amplitude between the tip and the substrate, $T$ is the temperature of the system, and $e$ is the elementary charge.

For simplicity we consider a conventional superconducting tip of the same material as the substrate, and, since the system is translation-invariant in the $y$-direction, we evaluate the LJC at $y=0$. In this case, combining Eq.~(\ref{eq:ljc}) with the Matsubara version of Eq.~(\ref{eq:F_local}) we are able to write the LJC as the sum of two terms:
\begin{equation}
I_{J}(x)=I_{e}(x) + I_{o}(x),
\label{eq:ljc_x}
\end{equation}    
where   
\begin{equation}
\begin{aligned}
I_{e}(x)=4et_0^2 \sin{\Phi} \ T \sum_n \int \frac{dk_y}{2\pi} &f_0(i\omega_n) F_{e}(x,k_y;i\omega_n), \\
I_{o}(x)=4et_0^2 \sin{\Phi} \ T \sum_n \int \frac{dk_y}{2\pi} &f_0(i\omega_n) F_{o}(x,k_y;i\omega_n). \\
\end{aligned}
\label{eq:LJC_eo}
\end{equation}
Here $F_{e/o}(i\omega_n)$ are the Matsubara versions of the expressions given in Eq.~(\ref{eq:feo}), while $f_0(i\omega_n)$ is the anomalous Green's function of the bare substrate given by:
\begin{equation}
f_0(i\omega_n)=\frac{-m\Delta}{2\pi\sqrt{\omega_n^2+\Delta^2}}\left[\arctan\left(\frac{\mu}{\sqrt{\omega_n^2+\Delta^2}} \right)+\frac{\pi}{2} \right].
\end{equation}
From Eq.~(\ref{eq:LJC_eo}) we find that $I_e$ is the contribution to the Josephson current coming strictly from even-$\omega$ pairing in the substrate, while $I_o$ represents the contribution to the current coming from the odd-$\omega$ pair amplitudes which are induced by the left nanowire and then reconverted to even-$\omega$ pairing by the right nanowire, $F_o$ in Eq.~(\ref{eq:F_local}). It is important to emphasize that this reconversion process is necessary to measure the odd-$\omega$ pair amplitudes because the LJC is not sensitive to the odd-$\omega$ odd-parity pair amplitudes present in this system, since those contributions vanish locally. We also note that, in real experiments it is the total LJC, Eq. (\ref{eq:ljc_x}), that is measured, not the separate contributions given by Eqs. (\ref{eq:LJC_eo}). However, it is instructive to examine these contributions separately since they allow us to discern which features in the total LJC are caused directly by the presence of odd-$\omega$ pairing and which are not. 

Given these expressions it is now straightforward to evaluate the separate contributions to the LJC for various cases similar to those presented in the previous sections. In principle, the LJC computed using Eq.~(\ref{eq:ljc_x}) is a function of the phase difference between the STM tip and the substrate. However, in practice, Josephson STM experiments probe only the maximum value of this quantity, given by setting $\Phi=\pi/2$, which is also what we report here. All results in this section were obtained using a temperature $T=\Delta/10$, below which we do not find significant changes to the LJC.

\begin{figure}
 \begin{center}
  \centering
        \includegraphics[width=0.45\textwidth]{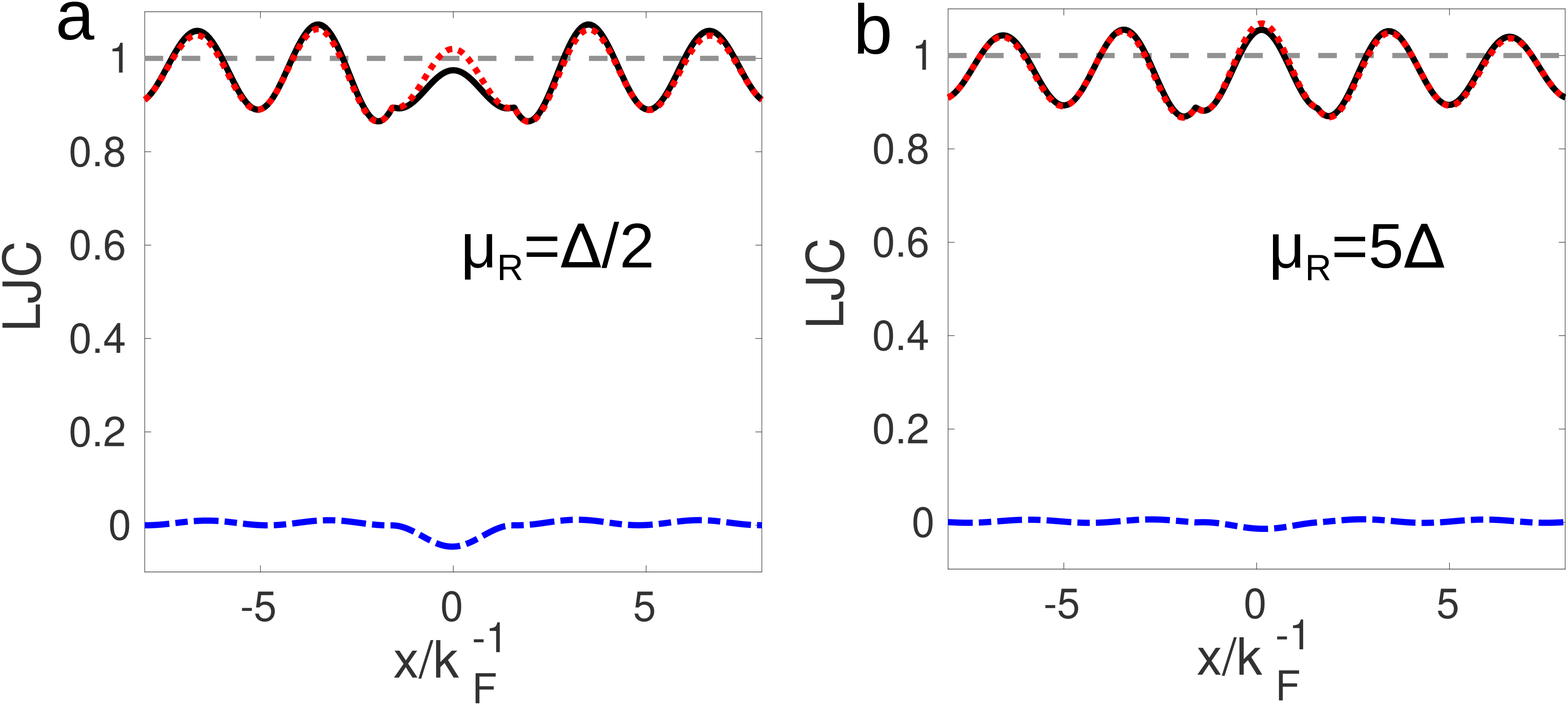}
  \caption{Maximum value of the LJC ($\Phi=\pi/2$) (solid/black) given by Eq.~(\ref{eq:ljc_x}) and separated into even-$\omega$ contributions (dotted/red) given by $I_e$ and odd-$\omega$ contributions (dashed/blue) given by $I_o$ in Eq.~(\ref{eq:LJC_eo}). The horizontal dashed line represents the LJC in the absence of the nanowires. We have fixed the model parameters such that $m_L=m_R=m_s$, $\mu_L=\Delta/2$, $\Delta=\mu_s/100$, $d=\pi k_F^{-1}$, where $k_F=\sqrt{2m_s\mu_s}$, and use two different values of $\mu_R$, as indicated in each panel.}
  \label{fig:LJC}
 \end{center}
\end{figure} 
In Fig.~\ref{fig:LJC} we plot the maximum LJC given by $I_J(x)$ in Eq.~(\ref{eq:ljc_x}), as well as its separate contributions, $I_e(x)$ and $I_o(x)$ given in Eq.~(\ref{eq:LJC_eo}), as a function of the position of the tip along the $x$-axis. We present these plots for fixed values of the effective masses, $m_R=m_L=m_s$, but two different values for the chemical potential in the right nanowire: $\mu_R=\Delta/2$ and $\mu_R=5\Delta$, to highlight the effect of the odd-$\omega$ pairing.

In both Figs.~\ref{fig:LJC} (a) and (b) we notice that the total LJC in the presence of the nanowires oscillates as a function of position $x$ around the value for the total LJC in the absence of the nanowires, and that this modulation decays with the distance from the nanowires, as expected. Further focusing on Fig.~\ref{fig:LJC} (a), the case in which both nanowire chemical potentials lie within the gap, $\mu_R=\mu_L=\Delta/2$, we notice a significant dip at $x=0$, directly between the two nanowires. Comparing the total value of the LJC to the components $I_e$ and $I_o$, we see that the total current is almost entirely composed of $I_e$, except in-between the two nanowires and especially at the peak position of the dip, $x=0$. In the region between the nanowires we instead also find a noticeable and negative contribution from the symmetry-converted odd-$\omega$ pair amplitudes, causing a suppression of the total LJC. Turning our attention to Fig.~\ref{fig:LJC} (b), in which only the left nanowire chemical potential lies within the gap, $\mu_R=5\Delta$, $\mu_L=\Delta/2$, we see that the dip in the LJC at $x=0$ has essentially vanished along with the contribution from the odd-$\omega$ pairing. 

\begin{figure}
 \begin{center}
  \centering
        \includegraphics[width=0.45\textwidth]{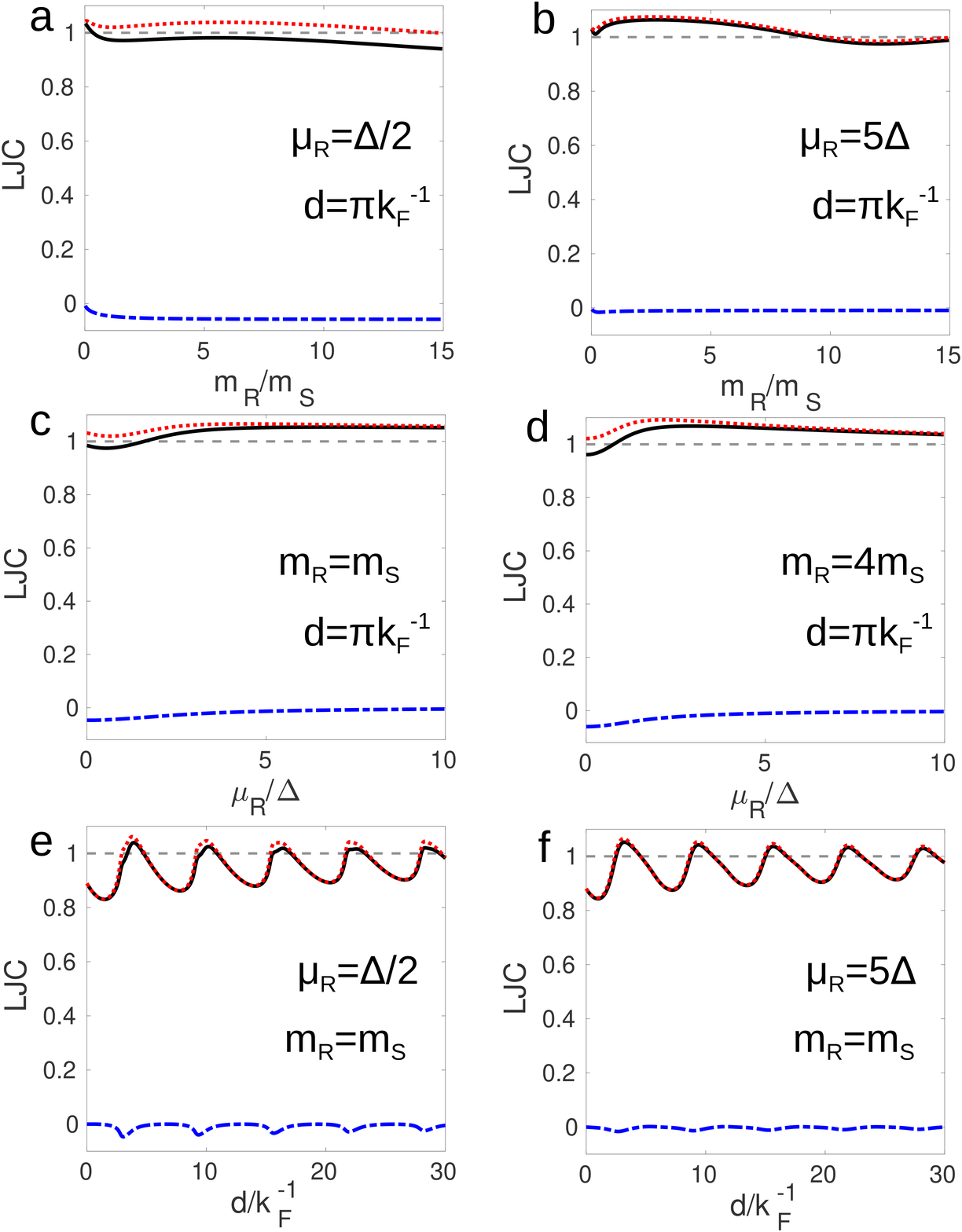}
  \caption{Maximum value of the LJC ($\Phi=\pi/2$) (solid/black) given by Eq.~(\ref{eq:ljc_x}) and separated into even-$\omega$ contributions (dotted/red) given by $I_e$ and odd-$\omega$ contributions (dashed/blue) given by $I_o$ in Eq.~(\ref{eq:LJC_eo}). The horizontal dashed line represents the LJC in the absence of the nanowires. We have fixed the position $x=0$ to focus on the suppression of $I_J$ relative to $I_e$, and we have set $m_L=m_s$, $\mu_L=\Delta/2$, $\Delta=\mu_s/100$, throughout, and present all distances in units of $k_F^{-1}=1/\sqrt{2m_s\mu_s}$. In each panel we plot these functions with respect to one of the model parameters: $m_R$, $\mu_R$, and $d$. In each panel the fixed values of the other two parameters are indicated.}
  \label{fig:LJC_cases}
 \end{center}
\end{figure} 

\begin{figure}
 \begin{center}
  \centering
        \includegraphics[width=0.45\textwidth]{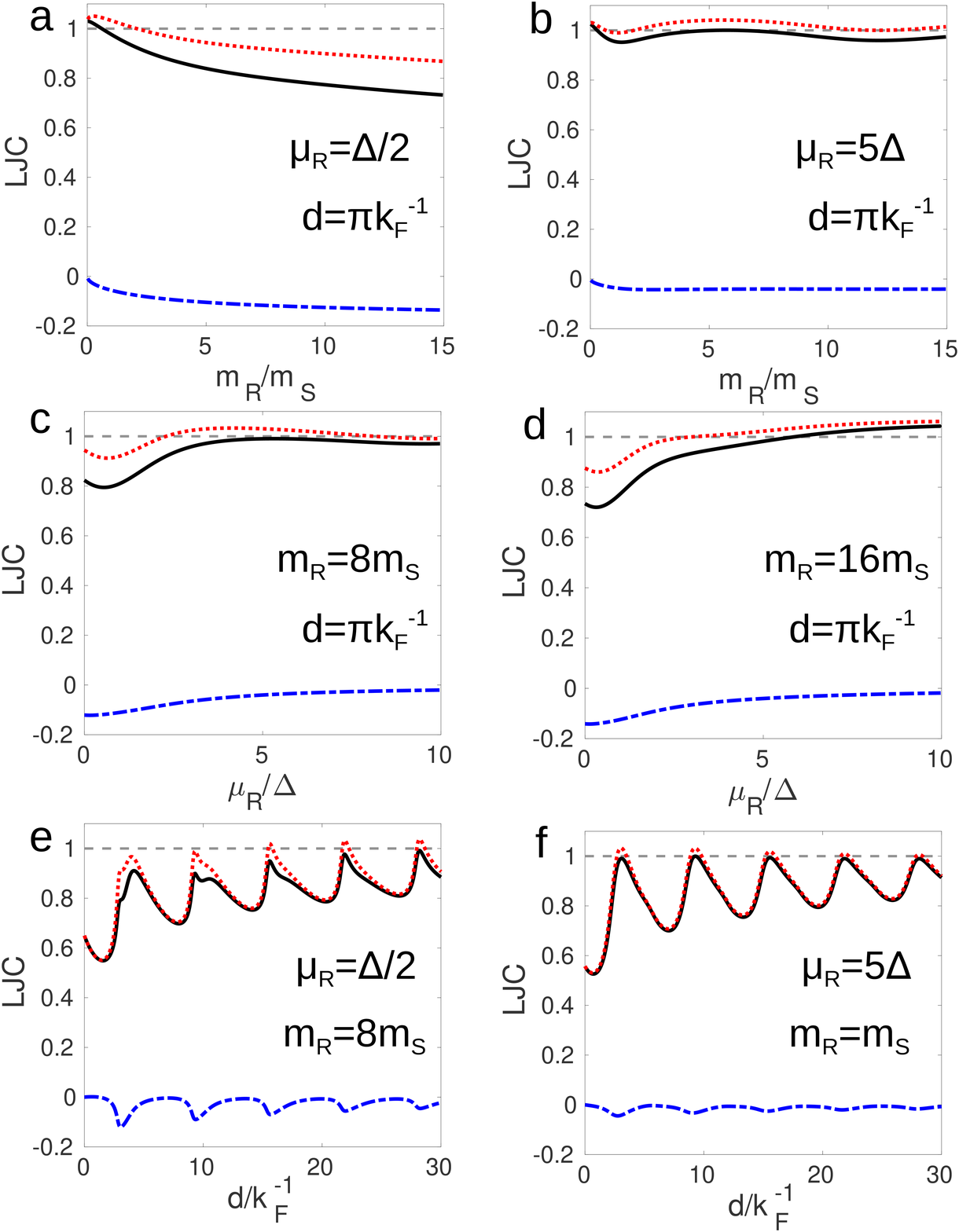}
  \caption{Maximum value of the LJC ($\Phi=\pi/2$) (solid/black) given by Eq.~(\ref{eq:ljc_x}) and separated into even-$\omega$ contributions (dotted/red) given by $I_e$ and odd-$\omega$ contributions (dashed/blue) given by $I_o$ in Eq.~(\ref{eq:LJC_eo}). The horizontal dashed line represents the LJC in the absence of the nanowires. We have fixed the position $x=0$ to focus on the suppression of $I_J$ relative to $I_e$, and, to highlight the effect of the odd-$\omega$ pairing, we have set $m_L=8m_s$, all other parameters are the same as Fig. \ref{fig:LJC_cases} except where specified. In each panel we plot these functions with respect to one of the model parameters: $m_R$, $\mu_R$, and $d$. The fixed values of the other two parameters are indicated.}
  \label{fig:LJC_cases_better}
 \end{center}
\end{figure} 

To further explore the behavior of the dip in the LJC we plot in Fig.~\ref{fig:LJC_cases} $I_J(x=0)$, $I_e(x=0)$, and $I_o(x=0)$ as functions of the chemical potential in the right nanowire, $\mu_R$, (a,b); the effective mass in the right nanowire, $m_R$, (c,d); and the interwire separation, $d$, (e,f), using the same parameters as in the plots of the pair amplitudes in Figs.~\ref{fig:L+SC+R_wmur}-\ref{fig:L+SC+R_wd}. In each case we find that the parameters which maximize the symmetry conversion of the odd-$\omega$ pairing also lead to a suppression of $I_J(x=0)$ relative to $I_e(x=0)$. Specifically, the suppression due to the odd-$\omega$ pairing is largest when $\mu_R \leq \Delta$, $d=(2n+1)\pi k_F^{-1}$, and $m_R>>m_s$. Moreover, this suppression can be turned off by increasing $\mu_R>>\Delta$ or setting $d$ away from $(2n+1)\pi k_F^{-1}$. While all these plots show an unambiguous effect of the odd-$\omega$ pair amplitudes, we note that the effect is somewhat small compared to the overall magnitude of the current. Therefore, in Fig.~\ref{fig:LJC_cases_better} we show the same plots as Fig.~\ref{fig:LJC_cases} but with larger values for the effective masses in the nanowires, to demonstrate that the effect of the odd-$\omega$ pairing can be considerably enhanced.     

In aggregate, the results in this section show that the presence of odd-$\omega$ odd-parity pair amplitudes directly influence local observables, both normal state properties, such as LDOS, and the superconducting Josephson effect, the LJC. These effects appear despite the fact that the odd spatial parity of the odd-$\omega$ pair amplitudes makes them intrinsically non-local. The reason the non-local odd-$\omega$ amplitudes influence measurable local properties is intrinsic to the double wire structure: non-local odd-$\omega$ pairing is induced in the substrate by coupling to one of the nanowires, then, higher-order tunneling processes between the substrate and the two nanowires results in the partial conversion of non-local odd-$\omega$ pairing to local even-$\omega$ pairing, which is then detectable by local probes. 

\section{Conclusions}
\label{sec:conclusions}

In this work we study the emergent properties of Cooper pair amplitudes in a system composed of two parallel nanowires separated by a distance $d$ and coupled to a conventional superconducting substrate with a spin-singlet $s$-wave order parameter. By expanding the anomalous Green's function perturbatively to leading order in the superconductor-nanowire tunneling amplitude, we show that odd-$\omega$ odd-parity Cooper pairing emerges in both the substrate and the interwire channel, despite the absence of spin-orbit coupling or magnetism in the system. We also provide simple analytic expressions to characterize the odd-$\omega$ pairing in each case.
Furthermore, by accounting for higher-order tunneling processes between the substrate and the two nanowires, we find that the non-local odd-$\omega$ pairing induced by the presence of one of the nanowires can be converted into local even-$\omega$ pairing by the presence of the other nanowire. We present semi-analytic results for the infinite-order pair amplitudes in terms of the odd-$\omega$ pairing and use these expressions to examine the conditions under which this higher-order symmetry conversion can be enhanced. 

Importantly, we use these results to study the effect of the odd-$\omega$ pair amplitudes on two local observables: the electronic local density of states (LDOS), measurable by scanning tunneling microscopy (STM), and the local Josephson current (LJC), measurable by Josephson STM. In the LDOS we find that certain subgap peaks obtain a contribution from the odd-$\omega$ pair amplitudes, and that the strength of this contribution depends strongly on the chemical potentials in the two nanowires. However, these subgap peaks also possess strong contributions from the even-$\omega$ pairing and normal quasiparticles, such that the total LDOS lacks strong features at these frequencies. In the LJC, we find that the odd-$\omega$ pair amplitudes provide a notable suppression of the maximum current in the region between the two nanowires. Moreover, this suppression can be tuned using various physical parameters, including the nanowire chemical potentials and effective masses, as well as the separation distance between the nanowires. Based on these results we predict that Josephson STM measurements are a particularly promising tool for studying the effects of odd-$\omega$ pairing directly, even when the odd-$\omega$ pairing is intrinsically odd in spatial parity and the STM tip possesses no odd-$\omega$ pairing itself. 

\acknowledgments 
We thank A.~V.~Balatsky, J.~Cayao, J.~Fransson, T.~L\"{o}thman, M.~Mashkoori, and F.~Parhizgar for useful discussions. This work was supported by the Swedish Research Council (Vetenskapsr\aa det) Grant No. 621-2014-3721, the Knut and Alice Wallenberg Foundation through the Wallenberg Academy Fellows program, and the European Research Council (ERC) under the European UnionÕs Horizon 2020 research and innovation programme (ERC-2017-StG-757553). We also acknowledge support from the COST Action CA16218 nanocohybri.\\
\\
\appendix 

\section{Green's functions for the substrate}
\label{app:g0}
In this appendix we present the exact expressions for the Matsubara and retarded Green's functions of the superconducting substrate in the absence of the two nanowires. Being a 2D homogenous superconductor, these are defined as:
\begin{equation}
\hat{\mathcal{G}}^{(0)}_{s}(x,k_y;z)=\int_{-\infty}^\infty\frac{dk_x}{2\pi}e^{ik_xx} \frac{z\hat{\tau}_0+\xi_{s,k}\hat{\tau}_3+\Delta\hat{\tau}_1}{z^2-\xi_{s,k}^2-\Delta^2},
\label{eq:gr_0}
\end{equation}
where the Matsubara Green's function is associated with imaginary frequencies, $z=i\omega_n$, while the retarded Green's function is associated with complex frequencies with poles only in the upper half plane, $z=\omega+i\eta$. In both cases, the Green's function takes the same form:
\begin{equation}
\hat{\mathcal{G}}^{(0)}_{s}(x,k_y;z)=\left[z\hat{\tau}_0+\Delta\hat{\tau}_1\right]g_0(x;z)+g_3(x;z)\hat{\tau}_3
\label{eq:gm_0_appendix}
\end{equation}
where the coefficients $g_0(x;z)$ and $g_3(x;z)$ may be found by converting the integral over $k_x$ in Eq.~(\ref{eq:gr_0}) to a contour integral over the counterclockwise-oriented contour covering the entire upper half-plane. However, care must be taken to ensure that all residues in the upper half-plane are accounted for. 

In the end, when $z=i\omega_n$, $g_0(x;z)$ and $g_3(x;z)$ are given by:
\begin{widetext}
\begin{equation}
\begin{aligned}
g_0(x;z=i\omega_n)&=-\frac{2m_s^2e^{-|x|k_0\sin\tfrac{\phi}{2}}}{k_0\Omega_n}\cos\left(|x|k_0\cos\tfrac{\phi}{2}-\frac{\phi}{2} \right) \\
g_3(x;z=i\omega_n)&= -\frac{m_se^{-|x|k_0\sin\tfrac{\phi}{2}}}{k_0\Omega_n}\left[k_0^2 \cos\left(|x|k_0\cos\tfrac{\phi}{2}+\frac{\phi}{2} \right)-\alpha \cos\left(|x|k_0\cos\tfrac{\phi}{2}-\frac{\phi}{2} \right) \right] \\\\
\end{aligned}
\label{eq:g0g3f_matsubara_appendix}
\end{equation}
\end{widetext} 
where $\Omega_n=2m_s\sqrt{\omega_n^2+\Delta^2}$, $\alpha=2m_s\mu_s-k_y^2$, $k_0=\left(\alpha^2+\Omega^2\right)^{1/4}$, and $\phi=\arctan(\Omega_n/\alpha)\in[0,\pi)$.
Whereas when $z=\omega+i\eta$, $g_0(x;z)$ and $g_3(x;z)$ are given by:
\begin{widetext}
\begin{equation}
\begin{aligned}
g_0(x;z=\omega+i\eta)&=\left\{\begin{array}{lc}
-\frac{m_s^2}{\tilde{\Omega}}\left( \frac{e^{i|x|\sqrt{\alpha+i\tilde{\Omega}}}}{\sqrt{\alpha+i\tilde{\Omega}}} +\frac{e^{-i|x|\sqrt{\alpha-i\tilde{\Omega}}}}{\sqrt{\alpha-i\tilde{\Omega}}} \right); & |\omega|<|\Delta| \\
-\frac{im_s^2}{\Omega}\left( \frac{e^{i|x|\sqrt{\alpha+\Omega}}}{\sqrt{\alpha+\Omega}} +\theta(\alpha-\text{Re}\Omega)\frac{e^{-i|x|\sqrt{\alpha-\Omega}}}{\sqrt{\alpha-\Omega}}+i\theta(\text{Re}\Omega-\alpha)\frac{e^{-|x|\sqrt{\Omega-\alpha}}}{\sqrt{\Omega-\alpha}} \right); & \omega>|\Delta| \\
\frac{im_s^2}{\Omega}\left( \frac{e^{i|x|\sqrt{\alpha-\Omega}}}{\sqrt{\alpha-\Omega}} +\theta(\alpha+\text{Re}\Omega)\frac{e^{-i|x|\sqrt{\alpha+\Omega}}}{\sqrt{\alpha+\Omega}}-\theta(-\alpha-\text{Re}\Omega)\frac{e^{i|x|\sqrt{\alpha+\Omega}}}{\sqrt{\alpha+\Omega}} \right); & \omega<-|\Delta| \\
\end{array} \right. \\
g_3(x;z=\omega+i\eta)&=\left\{\begin{array}{lc}
-\frac{im_s}{2}\left( \frac{e^{i|x|\sqrt{\alpha+i\tilde{\Omega}}}}{\sqrt{\alpha+i\tilde{\Omega}}} -\frac{e^{-i|x|\sqrt{\alpha-i\tilde{\Omega}}}}{\sqrt{\alpha-i\tilde{\Omega}}} \right); & |\omega|<|\Delta| \\
-\frac{im_s}{2}\left( \frac{e^{i|x|\sqrt{\alpha+\Omega}}}{\sqrt{\alpha+\Omega}} -\theta(\alpha-\text{Re}\Omega)\frac{e^{-i|x|\sqrt{\alpha-\Omega}}}{\sqrt{\alpha-\Omega}}-i\theta(\text{Re}\Omega-\alpha)\frac{e^{-|x|\sqrt{\Omega-\alpha}}}{\sqrt{\Omega-\alpha}} \right); & \omega>|\Delta| \\
-\frac{im_s}{2}\left( \frac{e^{i|x|\sqrt{\alpha-\Omega}}}{\sqrt{\alpha-\Omega}} -\theta(\alpha+\text{Re}\Omega)\frac{e^{-i|x|\sqrt{\alpha+\Omega}}}{\sqrt{\alpha+\Omega}}+\theta(-\alpha-\text{Re}\Omega)\frac{e^{i|x|\sqrt{\alpha+\Omega}}}{\sqrt{\alpha+\Omega}} \right); & \omega<-|\Delta| \\
\end{array} \right. \\
\end{aligned}
\label{eq:g0g3f_appendix}
\end{equation}
\end{widetext} 
Here $\theta$ is the Heaviside step function and we have defined $\Omega\equiv 2m_s\sqrt{(\omega+i\eta)^2-\Delta^2}$, and $\tilde{\Omega}\equiv 2m_s\sqrt{\Delta^2-(\omega+i\eta)^2}$.

\section{Coefficients in perturbative results}
By inserting the expressions from Eqs.~(\ref{eq:g0}) into Eqs.~(\ref{eq:delta_1}) we can explicitly calculate the leading order corrections to the anomalous Green's function of the superconducting substrate, presented in Eq.~(\ref{eq:dF1}), where, for compactness, we have defined the following three coefficients:
\begin{widetext}
\begin{equation}
\begin{aligned}
A_{x_1,x_2}(i\omega_n)=&\frac{t_L^2\exp\left[-\left(\left|x_1+\tfrac{d}{2}\right|+\left|x_2+\tfrac{d}{2}\right|\right)k_0\sin{\tfrac{\phi}{2}}\right]}{\omega_n^2+\xi_{L,k_y}^2} \sin\left[\left(\left|x_1+\tfrac{d}{2}\right|-\left|x_2+\tfrac{d}{2}\right|\right)k_0\cos{\tfrac{\phi}{2}}\right]  \\
+&\frac{t_R^2\exp\left[-\left(\left|x_1-\tfrac{d}{2}\right|+\left|x_2-\tfrac{d}{2}\right|\right)k_0\sin{\tfrac{\phi}{2}}\right]}{\omega_n^2+\xi_{R,k_y}^2} \sin\left[\left(\left|x_1-\tfrac{d}{2}\right|-\left|x_2-\tfrac{d}{2}\right|\right)k_0\cos{\tfrac{\phi}{2}}\right], 
\end{aligned}
\label{eq:A}
\end{equation}
\begin{equation}
\begin{aligned}
B_{x_1,x_2}(i\omega_n)=&\frac{t_L^2\exp\left[-\left(\left|x_1+\tfrac{d}{2}\right|+\left|x_2+\tfrac{d}{2}\right|\right)k_0\sin{\tfrac{\phi}{2}}\right]}{\omega_n^2+\xi_{L,k_y}^2} \left\{\cos\left[\left(\left|x_1+\tfrac{d}{2}\right|-\left|x_2+\tfrac{d}{2}\right|\right)k_0\cos{\tfrac{\phi}{2}}\right] \right. \\
&+\left.\cos\left[\left(\left|x_1+\tfrac{d}{2}\right|+\left|x_2+\tfrac{d}{2}\right|\right)k_0\cos{\tfrac{\phi}{2}}-\phi\right] \right\}  \\
+&\frac{t_R^2\exp\left[-\left(\left|x_1-\tfrac{d}{2}\right|+\left|x_2-\tfrac{d}{2}\right|\right)k_0\sin{\tfrac{\phi}{2}}\right]}{\omega_n^2+\xi_{R,k_y}^2} \left\{\cos\left[\left(\left|x_1-\tfrac{d}{2}\right|-\left|x_2-\tfrac{d}{2}\right|\right)k_0\cos{\tfrac{\phi}{2}}\right] \right. \\
&+\left.\cos\left[\left(\left|x_1-\tfrac{d}{2}\right|+\left|x_2-\tfrac{d}{2}\right|\right)k_0\cos{\tfrac{\phi}{2}}-\phi\right] \right\}, 
\end{aligned}
\label{eq:B}
\end{equation}
\begin{equation}
\begin{aligned}
C_{x_1,x_2}(i\omega_n)=&\frac{\xi_{L,k_y}t_L^2\exp\left[-\left(\left|x_1+\tfrac{d}{2}\right|+\left|x_2+\tfrac{d}{2}\right|\right)k_0\sin{\tfrac{\phi}{2}}\right]}{\omega_n^2+\xi_{L,k_y}^2} \sin\left[\left(\left|x_1+\tfrac{d}{2}\right|+\left|x_2+\tfrac{d}{2}\right|\right)k_0\cos{\tfrac{\phi}{2}}-\phi\right]  \\
+&\frac{\xi_{R,k_y}t_R^2\exp\left[-\left(\left|x_1-\tfrac{d}{2}\right|+\left|x_2-\tfrac{d}{2}\right|\right)k_0\sin{\tfrac{\phi}{2}}\right]}{\omega_n^2+\xi_{R,k_y}^2} \sin\left[\left(\left|x_1-\tfrac{d}{2}\right|+\left|x_2-\tfrac{d}{2}\right|\right)k_0\cos{\tfrac{\phi}{2}}-\phi\right], 
\end{aligned}
\label{eq:C}
\end{equation}
\end{widetext}
where $\Omega_n=2m_s\sqrt{\omega_n^2+\Delta^2}$, $\alpha=2m_s\mu_s-k_y^2$, $k_0=\left(\alpha^2+\Omega^2\right)^{1/4}$, and $\phi=\arctan(\Omega_n/\alpha)\in[0,\pi)$, as in Eq.~(\ref{eq:g0g3f_matsubara_appendix}). Clearly, all three functions in Eqs.~(\ref{eq:A})-(\ref{eq:C}) are even in Matsubara frequency $i\omega_n$ and $k_y$, since they depend on these variables through $\omega_n^2$ and $k_y^2$, respectively. Furthermore, we can see that $A_{x_1,x_2}=-A_{x_2,x_1}$, while $B_{x_1,x_2}=B_{x_2,x_1}$ and $C_{x_1,x_2}=C_{x_2,x_1}$.  

\section{Coefficients for $T$-Matrices and infinite-order Green's functions}
\label{app:tmatrix}
To study the influence of just the left nanowire on the superconducting substrate to infinite order in the tunneling, $t_L$, we use the $T$-matrix defined in Eq. (\ref{eq:T_L}). Using Eq.~(\ref{eq:schematic_g0}) together with the definition of $\hat{\Sigma}^{L}_s$ in Eq.~(\ref{eq:self}), we obtain an exact expression for $\hat{T}_L$, which takes the form given by Eq.~(\ref{eq:TL_explicit}) with the coefficients:

\begin{equation}
\begin{aligned}
T_0&=\frac{t_L^2(1-t_L^2g_0(0))}{z^2(1-t_L^2g_0(0))^2-(\xi_{L,k_y}+t_L^2g_3(0))^2-t_L^4\Delta^2g_0^2(0)} \\
T_1&=\frac{t_L^4\Delta g_0(0)}{z^2(1-t_L^2g_0(0))^2-(\xi_{L,k_y}+t_L^2g_3(0))^2-t_L^4\Delta^2g_0^2(0)} \\
T_3&=\frac{t_L^2(\xi_{L,k_y}+t_L^2g_3(0))}{z^2(1-t_L^2g_0(0))^2-(\xi_{L,k_y}+t_L^2g_3(0))^2-t_L^4\Delta^2g_0^2(0)}.
\end{aligned}
\label{eq:T013}
\end{equation}
While these are complicated expressions, we notice that these coefficients inherit the symmetries of $g_0$ and $g_3$, discussed in the main text. 
Furthermore, inserting Eq.~(\ref{eq:TL_explicit}) into Eq.~(\ref{eq:G_L}), we find the Green's function of the L+SC system to infinite order in the tunneling $t_L$, given by Eq. (\ref{eq:G_L_explicit}) with coefficients:
\begin{widetext}
\begin{equation}
\begin{aligned}
g^{(L)}_0(x_1,x_2)=&g_0(x_1-x_2)+T_0\left[\left(z^2+\Delta^2\right)g_0(x_1+\tfrac{d}{2})g_0(x_2+\tfrac{d}{2})+g_3(x_1+\tfrac{d}{2})g_3(x_2+\tfrac{d}{2}) \right] \\
&+T_3\left[ g_0(x_1+\tfrac{d}{2})g_3(x_2+\tfrac{d}{2})+g_3(x_1+\tfrac{d}{2})g_0(x_2+\tfrac{d}{2}) \right] +2\Delta T_1 g_0(x_1+\tfrac{d}{2})g_0(x_2+\tfrac{d}{2}), \\
g^{(L)}_3(x_1,x_2)=&g_3(x_1-x_2)+T_3\left[\left(z^2-\Delta^2\right)g_0(x_1+\tfrac{d}{2})g_0(x_2+\tfrac{d}{2})+g_3(x_1+\tfrac{d}{2})g_3(x_2+\tfrac{d}{2}) \right] \\
&+\left(z^2 T_0+\Delta T_1\right)\left[ g_0(x_1+\tfrac{d}{2})g_3(x_2+\tfrac{d}{2})+g_3(x_1+\tfrac{d}{2})g_0(x_2+\tfrac{d}{2}) \right], \\
f^{(L)}_1(x_1,x_2)=&\Delta g_0(x_1-x_2)+T_1\left[\left(z^2+\Delta^2\right)g_0(x_1+\tfrac{d}{2})g_0(x_2+\tfrac{d}{2})-g_3(x_1+\tfrac{d}{2})g_3(x_2+\tfrac{d}{2})\right] \\
&+ \Delta T_3\left[ g_0(x_1+\tfrac{d}{2})g_3(x_2+\tfrac{d}{2})+g_3(x_1+\tfrac{d}{2}) g_0(x_2+\tfrac{d}{2}) \right] +2\Delta z^2 T_0 g_0(x_1+\tfrac{d}{2})g_0(x_2+\tfrac{d}{2}), \\
f^{(L)}_2(x_1,x_2)=& \left(\Delta T_0+T_1 \right)\left[g_3(x_1+\tfrac{d}{2}) g_0(x_2+\tfrac{d}{2})- g_0(x_1+\tfrac{d}{2})g_3(x_2+\tfrac{d}{2})\right]. \\
\end{aligned}
\label{eq:GL0123}
\end{equation}
\end{widetext}

Next, turning our attention to the Green's function of the substrate in the presence of both nanowires, in Eq.~(\ref{eq:G_R}) we write this Green's function to infinite order in the tunneling, $t_R$, using the $T$-matrix defined in Eq.~(\ref{eq:T_R}). Inserting the expressions from Eqs.~(\ref{eq:self}) and (\ref{eq:localGL}) we find that this $T$-matrix takes the form given by Eq.~(\ref{eq:T_R_explicit}) where we define the coefficients:
\begin{widetext}
\begin{equation}
\begin{aligned}
T^{(R)}_0&=\frac{t_R^2(1-t_R^2g^{(L)}_0(\tfrac{d}{2},\tfrac{d}{2}))}{z^2(1-t_R^2g^{(L)}_0(\tfrac{d}{2},\tfrac{d}{2}))^2-(\xi_{R,k_y}+t_R^2g^{(L)}_3(\tfrac{d}{2},\tfrac{d}{2}))^2-t_R^4f^{(L)}_1(\tfrac{d}{2},\tfrac{d}{2})^2}, \\
T^{(R)}_1&=\frac{t_R^4f^{(L)}_1(\tfrac{d}{2},\tfrac{d}{2})}{z^2(1-t_R^2g^{(L)}_0(\tfrac{d}{2},\tfrac{d}{2}))^2-(\xi_{R,k_y}+t_R^2g^{(L)}_3(\tfrac{d}{2},\tfrac{d}{2}))^2-t_R^4f^{(L)}_1(\tfrac{d}{2},\tfrac{d}{2})^2}, \\
T^{(R)}_3&=\frac{t_R^2(\xi_{R,k_y}+t_R^2g^{(L)}_3(\tfrac{d}{2},\tfrac{d}{2}))}{z^2(1-t_R^2g^{(L)}_0(\tfrac{d}{2},\tfrac{d}{2}))^2-(\xi_{R,k_y}+t_R^2g^{(L)}_3(\tfrac{d}{2},\tfrac{d}{2}))^2-t_R^4f^{(L)}_1(\tfrac{d}{2},\tfrac{d}{2})^2}.
\end{aligned}
\label{eq:TR_123}
\end{equation}
\end{widetext}

Focusing on the local part of the Green's function, we find that Eq. (\ref{eq:G_R}) takes the form appearing in Eq. (\ref{eq:localGR}), where the coefficients $g^{(R)}_0$, $g^{(R)}_3$, and $f^{(R)}_1$ are: 
\begin{widetext}
\begin{equation}
\begin{aligned}
g^{(R)}_0(x)=&g^{(L)}_0(x,x)+T^{(R)}_0\left[z^2g^{(L)}_0(x,\tfrac{d}{2})^2+g^{(L)}_3(x,\tfrac{d}{2})^2+f^{(L)}_1(x,\tfrac{d}{2})^2+z^2f^{(L)}_2(x,\tfrac{d}{2})^2 \right] \\
&+2T^{(R)}_1\left[ g^{(L)}_0(x,\tfrac{d}{2})f^{(L)}_1(x,\tfrac{d}{2})+g^{(L)}_3(x,\tfrac{d}{2})f^{(L)}_2(x,\tfrac{d}{2}) \right] \\
&+2T^{(R)}_3\left[ g^{(L)}_0(x,\tfrac{d}{2})g^{(L)}_3(x,\tfrac{d}{2})-f^{(L)}_1(x,\tfrac{d}{2})f^{(L)}_2(x,\tfrac{d}{2}) \right], \\
g^{(R)}_3(x)=&g^{(L)}_3(x,x)+T^{(R)}_3\left[z^2g^{(L)}_0(x,\tfrac{d}{2})^2+g^{(L)}_3(x,\tfrac{d}{2})^2-f^{(L)}_1(x,\tfrac{d}{2})^2-z^2f^{(L)}_2(x,\tfrac{d}{2})^2 \right] \\
&+2z^2T^{(R)}_0\left[ g^{(L)}_0(x,\tfrac{d}{2})g^{(L)}_3(x,\tfrac{d}{2})+f^{(L)}_1(x,\tfrac{d}{2})f^{(L)}_2(x,\tfrac{d}{2}) \right] \\
&+2T^{(R)}_1\left[z^2 g^{(L)}_0(x,\tfrac{d}{2})f^{(L)}_2(x,\tfrac{d}{2})+g^{(L)}_3(x,\tfrac{d}{2})f^{(L)}_1(x,\tfrac{d}{2}) \right], \\
f^{(R)}_1(x)=&f^{(L)}_1(x,x)+T^{(R)}_1\left[z^2g^{(L)}_0(x,\tfrac{d}{2})^2-g^{(L)}_3(x,\tfrac{d}{2})^2+f^{(L)}_1(x,\tfrac{d}{2})^2-z^2f^{(L)}_2(x,\tfrac{d}{2})^2 \right] \\
&+2z^2T^{(R)}_0\left[ g^{(L)}_0(x,\tfrac{d}{2})f^{(L)}_1(x,\tfrac{d}{2})-g^{(L)}_3(x,\tfrac{d}{2})f^{(L)}_2(x,\tfrac{d}{2}) \right] \\
&-2T^{(R)}_3\left[z^2 g^{(L)}_0(x,\tfrac{d}{2})f^{(L)}_2(x,\tfrac{d}{2})-g^{(L)}_3(x,\tfrac{d}{2})f^{(L)}_1(x,\tfrac{d}{2}) \right], \\
\end{aligned}
\label{eq:GR0123}
\end{equation}
where we have repeatedly made use of the definitions in Eqs.~(\ref{eq:GL0123}) and (\ref{eq:TR_123}).
\end{widetext}

\bibliographystyle{apsrevmy}
\bibliography{Odd_Frequency_Nanowires}

\end{document}